\documentclass[11pt]{article}
\usepackage[margin=2cm,right=3cm,left=3cm]{geometry}

\usepackage{amsmath,amsthm,amsfonts,dsfont}
\usepackage{xcolor}
\usepackage{graphicx}
\usepackage{nicefrac}
\usepackage{xspace}
\usepackage{booktabs}
\usepackage{caption}
\usepackage{subcaption}
\usepackage{aligned-overset}
\usepackage{enumerate}
\usepackage{cite}
\usepackage{tikz}
\usetikzlibrary{automata, positioning}
\usepackage{makecell} % supports line breaks inside cells
\usepackage{hyperref}
\usepackage[ruled,linesnumbered]{algorithm2e}
\usepackage{cleveref}

\SetAlCapSkip{8pt}

\DeclareMathOperator{\OPT}{OPT}

\newtheorem{theorem}{Theorem}
\numberwithin{theorem}{section}
\newtheorem{lemma}[theorem]{Lemma}
\newtheorem{definition}[theorem]{Definition}

\newcommand{\NN}{\mathds{N}}

\newcommand{\card}[1]{\left\vert{#1}\right\vert }

\renewcommand{\Pr}[1]{\mbox{\rm\bf Pr}\left[#1\right]}
\newcommand{\E}[1]{\mbox{\rm\bf E}\left[#1\right]}

\DeclareMathOperator{\pos}{pos}
\newcommand{\intD}[1]{~\mathrm{d} {#1}~}

\newtheorem{fact}{Fact}

\title{New Results for the \texorpdfstring{$k$}{k}-Secretary Problem\thanks{
Work supported by the European Research Council, Grant Agreement No.\ 691672. An earlier version of this paper has appeared in \textit{Proceedings of 30th International Symposium on Algorithms and Computation (ISAAC 2019)}.}}

\author{Susanne Albers \\
	Department of Computer Science, \\Technial University of Munich,\\
	albers@in.tum.de \and 
	Leon Ladewig\\
	Department of Computer Science, \\Technial University of Munich,\\
	ladewig@in.tum.de
}

\date{}

\newcommand{\RR}{\mathds{R}}

\renewcommand{\Pr}[1]{\mbox{\rm\bf Pr}\left[#1\right]}
\newcommand{\fallingfactorial}[2]{#1^{\underline{#2}}}
\newcommand{\intInter}[2]{\ensuremath{[#1..#2]}}
\newcommand{\intInterLeftOpen}[2]{\ensuremath{(#1..#2]}}
\newcommand{\intInterRightOpen}[2]{\ensuremath{[#1..#2)}}
\newcommand{\intInterOpen}[2]{\ensuremath{(#1..#2)}}

\begin{document}
	
\maketitle
\allowdisplaybreaks

\begin{abstract}
Suppose that $n$ items arrive online in random order and the goal is to select $k$ of them
such that the expected sum of the selected items is maximized.
The decision for any item is irrevocable and must be made on arrival without knowing future items.
This problem is known as the \textit{$k$-secretary problem}, which includes the classical secretary problem with the special case 
$k=1$. It is well-known that the latter problem can be solved by a simple algorithm of competitive ratio $1/e$ which is optimal for $n \to \infty$.
Existing algorithms beating the threshold of $1/e$ either rely on involved selection policies already for $k=2$, 
or assume that $k$ is large.

In this paper we present results for the $k$-secretary problem, considering the interesting and relevant case that $k$ is small. We focus on simple selection algorithms, accompanied by combinatorial analyses.
As a main contribution we propose a natural deterministic algorithm designed to have competitive ratios strictly greater than $1/e$ for small $k \geq 2$. This algorithm is hardly more complex than the elegant strategy for the classical secretary problem, optimal for $k=1$, and works for all $k \geq 1$. We derive its competitive ratios for $k \leq 100$, ranging from $0.41$ for $k=2$ to $0.75$ for $k=100$.

Moreover, we consider an algorithm proposed earlier in the literature, for which no rigorous analysis is known.
We show that its competitive ratio is $0.4168$ for $k=2$, implying that the previous analysis was not tight. Our analysis reveals a surprising combinatorial property of this algorithm, which might be helpful to find a tight analysis for all $k$.
\end{abstract}

\section{Introduction}
The \textit{secretary problem} is a well-known problem in the field of optimal stopping theory and is defined as follows:
Given a sequence of $n$ items which arrive online and in random order, select the maximum item.
The decision to accept or reject an item must be made immediately and irrevocably upon its arrival, especially without knowing future items.
The statement of the problem dates back to the 1960s and the optimal algorithm was published by Lindley \cite{lindley1961dynamic} and Dynkin \cite{dynkin1963optimum}.
For discussions on the origin of the problem, we refer to the survey by Ferguson \cite{ferguson89survey}.

In the past years, generalizations of the secretary problem involving selection of multiple items have become very popular.
We consider one of the most canonical generalizations known as the \textit{$k$-secretary problem}: Here, the algorithm is allowed to choose $k$ elements and the goal
is to maximize the expected sum of accepted elements. Other objective functions, such as maximizing the
probability of accepting the $k$ best \cite{freeman1983secretary,ajtai2001improved}
or general submodular functions \cite{DBLP:conf/approx/KesselheimT17}, have been studied as well.
Maximizing the sum of accepted items is closely related to the \textit{knapsack secretary problem} \cite{DBLP:conf/stoc/KesselheimTRV14,DBLP:conf/approx/BabaioffIKK07,DBLP:conf/approx/Albers0L19}.
If all items have unit weight and thus the capacity constraint is a cardinality bound, the $k$-secretary
problem arises.
The \textit{matroid secretary problem}, introduced by Babaioff et al.\ \cite{DBLP:conf/soda/BabaioffIK07},
is a generalization where an algorithm must maintain a set of accepted items that form an independent set of a given matroid.
We refer the reader to \cite{DBLP:conf/focs/Lachish14,DBLP:conf/soda/FeldmanSZ15,DBLP:conf/soda/FeldmanSZ18}
for recent work. If the matroid is $k$-uniform, again, the $k$-secretary problem occurs.
Another closely related problem was introduced by 
Buchbinder, Jain, and Singh \cite{DBLP:journals/mor/BuchbinderJS14}.
In the \textit{$(J,K)$-secretary problem}, an algorithm has $J$ choices and the objective is to maximize the number of selected items among the $K$ best. 
It can be shown that any monotone algorithm for the $(k,k)$-secretary problem corresponds to a $k$-secretary algorithm of the same competitive ratio \cite{DBLP:journals/mor/BuchbinderJS14}. Here, an algorithm is monotone if for any pair of items, it accepts the better item with higher probability. 
On the other side, any ordinal algorithm for $k$-secretary can be transformed to an algorithm for the $(k,k)$-secretary problem while maintaining the competitive ratio \cite{DBLP:journals/mor/BuchbinderJS14}.
Ordinal algorithms \cite{DBLP:conf/icalp/HoeferK17} decide based on the total order of items only, rather than on their numeric values. In fact, most known and elegant algorithms for the $k$-secretary problem are ordinal \cite{lindley1961dynamic,dynkin1963optimum,DBLP:conf/approx/BabaioffIKK07,DBLP:conf/soda/Kleinberg05}.

The large interest in generalizations of the classical secretary problem is motivated mainly by
numerous applications in online market design 
\cite{DBLP:conf/soda/BabaioffIK07,DBLP:conf/soda/Kleinberg05,DBLP:journals/sigecom/BabaioffIKK08}.
Apart from these applications, the secretary problem is the prototype of an online problem
analyzed in the  random order model:
An adversarial input order often rules out good (or even constant) competitive ratios when considering online optimization 
problems without further constraints.
By contrast, the assumption that the input is ordered randomly improves the competitive ratios in many optimization problems. This includes packing problems \cite{DBLP:conf/stoc/KesselheimTRV14,DBLP:conf/approx/Albers0L19,DBLP:conf/soda/Kenyon96}, scheduling problems \cite{DBLP:conf/esa/0002KT15}, and graph problems \cite{DBLP:conf/stoc/MahdianY11,DBLP:conf/soda/BahmaniMM10}.
Therefore, developing new techniques for secretary problems may, more generally, yield relevant insights for this input model as well.

\subsection{Previous Work}
The $k$-secretary problem was introduced by Kleinberg \cite{DBLP:conf/soda/Kleinberg05} in 2005.
He presents a randomized algorithm attaining a competitive ratio of $1 - 5 / \sqrt{k}$, which approaches 
$1$ for $k \to \infty$.
Moreover, Kleinberg shows that any algorithm has a competitive ratio of $1 - \Omega (\sqrt{1/k})$.
Therefore, from an asymptotic point of view, the $k$-secretary problem is solved by Kleinberg's result.
However, the main drawback can be seen in the fact that the competitive ratio is not
defined for $k \leq 24$ and breaks the barrier of $1/e$ only if $k \geq 63$ 
(see \Cref{fig:singlerefCompetitiveRatiosFull}, p.~\pageref{fig:singlerefCompetitiveRatiosFull}).

In 2007 the problem was revisited by Babaioff et al.\ \cite{DBLP:conf/approx/BabaioffIKK07}.
The authors propose two algorithms called \textsc{virtual} and \textsc{optimistic} and prove that both
algorithms have a competitive ratio of at least $1/e$ for any $k$. While the analysis of \textsc{virtual} is simple and tight, it takes much more
effort to analyze \textsc{optimistic} \cite{DBLP:conf/approx/BabaioffIKK07,DBLP:journals/sigecom/BabaioffIKK08}.
The authors believe that their analysis for \textsc{optimistic} is not tight for $k \geq 2$.

Further indications for competitive ratios strictly greater than $1/e$ can be obtained from the framework of
Buchbinder, Jain, and Singh \cite{DBLP:journals/mor/BuchbinderJS14}. In this framework, optimal algorithms for $(J,K)$-secretary and other variants of the secretary problem can be obtained using linear programming techniques.
By numerical simulations for the $(k,k)$-secretary problem with $n=100$, Buchbinder et al.\ obtained competitive ratios of $0.474$, $0.565$, and $0.612$, for $k=2,3,$ and $4$, respectively.
However, deriving an algorithm from their framework requires a formal analysis of the corresponding LP in the limit of $n \to \infty$,
which is not provided in the article \cite[p.~192]{DBLP:journals/mor/BuchbinderJS14}.

Chan, Chen, and Jiang \cite{DBLP:conf/soda/ChanCJ15} revisited the $(J,K)$-secretary problem and obtained several fundamental results. Notably, they showed that optimal algorithms for the $k$-secretary problem require access to the numeric values of the items, which complements the previous line of research in the ordinal model.
Chan et al.\ demonstrate this by providing a $0.4920$-competitive algorithm for the 2-secretary problem which is based on an algorithm for the $(2,2)$-secretary problem of competitive ratio $0.4886$.
Still, a rigorous analysis for the general $(J,K)$-secretary problem revealing the numeric competitive ratios is not known, even for $J=K$. 
Moreover, the resulting algorithms seem overly involved.
This dims the prospect of elegant $k$-secretary algorithms for $k \geq 3$ obtained from this approach.

\subsection{Our Contribution}

We study the $k$-secretary problem, the most natural and immediate generalization of the classical secretary problem.
While the extreme cases $k=1$ and $k \to \infty$ are well studied, hardly any results for small values of $k \geq 2$ exist. 
We believe that simple selection algorithms, performing well for small $k$, are interesting both from a theoretical point of view and for practical settings. Moreover, the hope is that existing algorithms for related problems based on $k$-secretary algorithms can be improved this way \cite[p.~191]{DBLP:journals/mor/BuchbinderJS14}.
We study ordinal, threshold-based algorithms in the style of \cite{lindley1961dynamic,dynkin1963optimum}.

As main contribution, we propose and analyze a simple deterministic algorithm \textsc{single-ref}.
This algorithm uses a single reference value as threshold for accepting items. To the best of our knowledge, this approach has not been explored for the $k$-secretary problem so far, although this natural idea arises in algorithms for related problems \cite{DBLP:journals/ior/AgrawalWY14}.
As a strength of our algorithm we see its simplicity: It is of plain combinatorial nature and can be 
fine-tuned using only two parameters. In contrast, the optimal algorithms following theoretically from the $(J,K)$-secretary approach \cite{DBLP:conf/soda/ChanCJ15} involve $k^2$ parameters and the same number of different decision rules.

The analysis of \textsc{single-ref} crucially depends on the fact that items can be partitioned into two classes, which we will call \textit{dominating} and \textit{non-dominating}. Both have certain properties on which we base our fully parameterized analysis.
In \Cref{tab:comeptitiveRatiosIntro}, we list the competitive ratios of \textsc{single-ref} for $k \leq 20$ assuming $n \to \infty$.
While the competitive ratio for $k=1$ is optimal, we obtain a value significantly greater than $1/e$ already for $k=2$. Furthermore, the competitive ratios are monotonically increasing in the interval $k \in \intInter{1}{20}$, 
already breaking the threshold of $0.5$ at $k=6$. Numerical computations suggest that this monotonicity 
holds for general $k$.
See \Cref{fig:singlerefCompetitiveRatiosFull} (p.~\pageref{fig:singlerefCompetitiveRatiosFull}) for the competitive ratios up to $k=100$ and a comparison with Kleinberg's algorithm  \cite{DBLP:conf/soda/Kleinberg05}. Providing a closed formula for the competitive ratio for any value of $k$ is one direction of future work (see \Cref{sec:conclusion}).

Moreover, we investigate the \textsc{optimistic} algorithm by Babaioff et al.\ \cite{DBLP:conf/approx/BabaioffIKK07} for the case $k=2$. Although Chan et al.\ \cite{DBLP:conf/soda/ChanCJ15} provided a strong algorithm for $k=2$, we think studying this elegant algorithm is interesting for two reasons:
First, a tight analysis of \textsc{optimistic} is stated as open problem in \cite{DBLP:conf/approx/BabaioffIKK07}.
Article \cite{DBLP:conf/approx/BabaioffIKK07} does not provide the proof of the $(1/e)$-bound and a recent journal publication \cite{DBLP:journals/jacm/BabaioffIKK18} (evolved from \cite{DBLP:conf/approx/BabaioffIKK07} and \cite{DBLP:conf/soda/BabaioffIK07}) does not cover the \textsc{optimistic} algorithm at all.
We make progress in this problem by proving that its competitive ratio is exactly $0.4168$ for $k=2$, which significantly breaks the $(1/e)$-barrier.
Second, our proof reveals an interesting property of this algorithm, which we show in \Cref{lemma:optimisticP2}: The probability that \textsc{optimistic} accepts the second best item is exactly the probability
that the optimal algorithm for $k=1$ accepts the best item.
A similar property might hold for $k \geq 3$, which could be a key insight into the general case.

\begin{table}
	\centering
	\begin{small}
	\begin{tabular}{ccccccccccc} 
		$k$ & 1 & 2 & 3 & 4 & 5 & 6 & 7 & 8 & 9 & 10 \\
		\midrule
		$\alpha$
		& $1/e$
		& 0.41
		& 0.44
		& 0.47 
		& 0.49 
		& 0.51 
		& 0.53 
		& 0.54 
		& 0.55 
		& 0.56 
		~\\[1em]
		$k$ & 11 & 12 & 13 & 14 & 15 & 16 & 17 & 18 & 19 & 20 \\
		\midrule
		$\alpha$
		& 0.57
		& 0.58
		& 0.59
		& 0.59 
		& 0.60
		& 0.60
		& 0.61
		& 0.62
		& 0.62
		& 0.63
	\end{tabular}
	\end{small}
\caption{Competitive ratios $\alpha$ of \textsc{single-ref} for $k \in \intInter{1}{20}$.}
\label{tab:comeptitiveRatiosIntro}		
\end{table}

From a technical point of view, we analyze the algorithms using basic combinatorial constructs exclusively.
This is in contrast to previous approaches \cite{DBLP:journals/mor/BuchbinderJS14,DBLP:conf/soda/ChanCJ15} which can only be analyzed using heavyweight linear programming techniques.
The combinatorial parts of our analysis are exact and hold for all $n$. In order to evaluate the competitive ratios numerically, we find lower bounds that hold for sufficiently large $n$.
Throughout the analyses of both algorithms, we associate probabilities with sets of permutations (see \Cref{sec:randomArrivalOrder}). Hence, probability relations can be shown equivalently by set relations.
This is a simple but powerful technique which may be useful in the analysis of other optimization problems with random arrival order as well.

\section{Preliminaries}
\label{sec:prelim}
Let $v_1 > v_2 > \ldots > v_n$ be the \textit{elements} (also called \textit{items}) of the input.
Note that we can assume w.l.o.g.\ that all items are distinct in the ordinal model.
Therefore, we say that $i$ is the \textit{rank} of element~$v_i$.
An \textit{input sequence} is any permutation of the list $v_1,\ldots,v_n$.
We denote the position of an element $v$ in the input sequence $\pi$ with $\pos_\pi(v) \in \{1,\ldots,n\}$
and write $\pos(v)$ whenever the input sequence is clear from the context.

Given any input sequence, an algorithm can accept up to $k$ items, where the decision
whether to accept or reject an item must be made immediately upon its arrival.
Let $\mathrm{ALG}$ denote the sum of items accepted by the algorithm. The algorithm is \textit{$\alpha$-competitive}
if $\E{\mathrm{ALG}} \geq \left( \alpha - o(1) \right) \cdot \OPT$ holds for all item sets, where
the expectation is taken over the uniform distribution of all $n!$ input sequences. 
Throughout the paper, $o(1)$ terms are asymptotic with respect to the number of items $n$ and
$\OPT = \sum_{i=1}^{k} v_i$ denotes the value of an optimal offline solution.

\subsection{Notation}
For $a, b \in \NN$ with $a \leq b$, we 
use the notation $\intInter{a}{b}$ to denote the set of integers $\{a,a+1,\ldots,b\}$ 
and write $[a]$ for $\intInter{1}{a}$.
The (half-)open integer intervals $\intInterLeftOpen{a}{b}$, $\intInterRightOpen{a}{b}$, and $\intInterOpen{a}{b}$
are defined accordingly.
Further, we use the notation $\fallingfactorial{n}{k}$ for the falling factorial $\frac{n!}{(n-k)!}$.

\subsection{Random Permutations}
\label{sec:randomArrivalOrder}
We often use the following process to obtain a permutation drawn uniformly at random:
Fix any order $u_1, u_2, \ldots,u_n$ of positions. Then, draw the element for position $u_1$ uniformly at random among all $n$ elements, next the element for position $u_2$ among the remaining $n-1$ elements, and so on.

Moreover, the uniform distribution of permutations allows us to prove probability relations using functions:
Suppose that $p_i$ is the probability that item $v_i$ is accepted in a random permutation.
Then $p_i = \card{P_i} / n!$, where $P_i$ is the set of all input sequences where $v_i$ is accepted.
Thus, we can prove $p_i \leq p_j$ by finding an injective function $f \colon P_i \to P_j$
and get $p_i = p_j$ if $f$ is bijective.
This technique turns out to be highly useful (e.g.\ in the proof of \Cref{lemma:optimisticP2}, which relates probabilities of different algorithms).

\subsection{Combinatorics}
\label{sec:combinatorics}

When analyzing random permutations, we often need to analyze the probability that $K$ items in a sequence have a certain property. This is described by the following random experiment which is a special case of the hypergeometric distribution.

\begin{fact}
	\label{fact:BlueBalls}
	Suppose there are $N$ balls in an urn from which $M$ are blue and $N-M$ red.
	The probability of drawing $K$ blue balls without replacement in a sequence of length $K$ is 
	$ h(N,M,K) := \binom{M}{K} / \binom{N}{K} \,.$
\end{fact}
\noindent

To simplify the binomial coefficients arising from $h(N,M,K)$, we make use of three useful identities (see \cite{graham1994concrete}) stated in the following.

For integers $l,m,n,w$ with $l,m \geq 0$ and $n \geq w \geq 0$, it holds that
\begin{equation}
\label{rule:sumOfProducts}
\tag{R1}
\sum_{k=0}^{l} \binom{l-k}{m} \binom{w+k}{n} = \binom{l+w+1}{m+n+1} \,.
\end{equation}
The well-known symmetry property for binomial coefficients states that for any integers
$n$, $k$ with $n \geq 0$,
\begin{equation}
\label{rule:symmetry}
\tag{R2}
\displaystyle{\binom{n}{k} = \binom{n}{n-k}} \,.
\end{equation}
Finally, by the trinomial revision rule, for any integers $m$ and $k$, and any real number $r$, it holds that
\begin{equation}
\label{rule:trinomialRevision}
\tag{R3}
\displaystyle{\binom{r}{m} \binom{m}{k} = \binom{r}{k} \binom{r-k}{m-k}} \,.
\end{equation}

\subsection{Bounding Sums by Integrals}
To bound sums over monotone increasing summands we make use of the following facts.

\begin{fact}
	\label{lemma:approxSumByIntegral}
	Let $f \colon \RR_{\geq 0} \to \RR_{\geq 0} $ and $a,b \in \NN$. If $f$ is
	\begin{enumerate}[(A)]
		\item
		monotonically decreasing, then 
		$	\int_{a}^{b+1} f(i) \intD{i} \leq \sum_{i=a}^{b} f(i) \leq \int_{a-1}^{b} f(i) \intD{i}$\!.
		\label{item:approxSumByIntegralDecreasing}
		\item monotonically increasing, then 
		$	\int_{a-1}^{b} f(i) \intD{i} \leq \sum_{i=a}^{b} f(i) \leq \int_{a}^{b+1} f(i) \intD{i}$\!.
		\label{item:approxSumByIntegralIncreasing}	
	\end{enumerate}
\end{fact}

\section{Algorithms}
\label{sec:algorithms}

In this section, we state our proposed algorithm \textsc{single-ref} and the \textsc{optimistic} algorithm by Babaioff et al.\ \cite{DBLP:conf/approx/BabaioffIKK07} and compare both strategies.
While both algorithms have an initial sampling phase in which the first $t-1$ items are rejected, 
the main difference is the policy for accepting items:
Let $s_j$ be the $j$-th best item from the sampling.

\begin{algorithm}
	\SetAlgoLined
	\DontPrintSemicolon
	\SetKwInOut{Parameter}{Parameters}
	
	\Parameter{$t \in \intInterLeftOpen{k}{n-k}$ (sampling threshold), \\
		$r \in [k]$ (reference rank)}
	
	\textbf{Sampling phase}: Reject the first $t-1$ items. \;
	Let $s_r$ be the $r$-th best item from the sampling phase. \;
	\textbf{Selection phase:} Choose the first $k$ items better than $s_r$. \;
	
	\caption{\textsc{single-ref}}
	\label[algorithm]{alg:simple}
\end{algorithm}
\begin{algorithm}
	\SetAlgoLined
	\DontPrintSemicolon
	\SetKwInOut{Parameter}{Parameters}
	
	\Parameter{$t \in \intInterLeftOpen{k}{n-k}$ (sampling threshold)}
	
	\textbf{Sampling phase}: Reject the first $t-1$ items. \;
	Let $s_1 > \ldots > s_k$ be the $k$ best items from the sampling phase. \;
	\textbf{Selection phase:} As the $j$-th accepted item, choose the first item better than $s_{k-j+1}$. \;
	
	\caption{\textsc{optimistic} \cite{DBLP:conf/approx/BabaioffIKK07}}
	\label[algorithm]{alg:optimistic}
\end{algorithm}
			
	\textsc{single-ref} uses only item $s_r$ as reference element.
	In the selection phase, the algorithm accepts the first $k$ elements better than $s_r$.
	Despite its simple structure, a challenging part in the analysis of \textsc{single-ref} is the dependence between both parameters $r$ and $t$.
	
	\textsc{optimistic} uses the $k$ best items from the sampling as reference elements.
	Right after the sampling phase, the first item better than $s_k$ will be accepted.
	The following accepted items are chosen similarly, but with $s_{k-1}, s_{k-2}, \ldots, s_1$ as reference items.
	Note that \textsc{optimistic} sticks to this order of reference elements, even if the first item already
	outperforms $s_1$.	Hence, it is optimistic in the sense that it always expects high-value items in the future.
	
	Note that in the case $k=1$, both \textsc{optimistic} and \textsc{single-ref} become the optimal algorithm for the secretary problem \cite{lindley1961dynamic,dynkin1963optimum}:
	After rejecting the first $t-1$ items, choose the first one better than the best from sampling.
	This strategy selects the best item with probability 
	$\frac{t-1}{n}\sum_{i=t}^{n} \frac{1}{i-1}$. 
	
	From now on, let
	\[
	p_i := \Pr{\mathcal{A} \text{ accepts item $v_i$}} \,,
	\]
	where $\mathcal{A}$ is either $\textsc{single-ref}$ or $\textsc{optimistic}$ and $i \in [n]$.
	We further define
	\[
	p_i^{(j)} := \Pr{\mathcal{A} \text{ accepts item $v_i$ as the $j$-th item}}
	\]
	for $i \in [n]$ and $j \in [k]$. Clearly, $p_i = \sum_{j=1}^{k} p_i^{(j)}$.
	Next, we define monotonicity of $k$-secretary algorithms and prove this property for both algorithms.
	\begin{definition}
		An algorithm is called \textit{monotone} if $p_i \geq p_j$ holds for any two items $v_i > v_j$.
	\end{definition}
	
	\begin{lemma}
	\label{lemma:monotonicity}
	\textsc{optimistic} and \textsc{single-ref} are monotone.
	\end{lemma}
	\begin{proof}
		We prove that $p_i \geq p_{i+1}$ for all $i \in [n-1]$.
		By the concept described in \Cref{sec:randomArrivalOrder}, it is sufficient to show that for each input sequence where $v_{i+1}$ is accepted, there exists a unique input sequence where $v_i$ is accepted.
		
		Consider any input sequence $\pi$ in which $v_{i+1}$
		is accepted. Let $s_j < v_{i+1}$ be the sampling item to which $v_{i+1}$ is compared (in case of \textsc{single-ref}, we have $j=r$).
		Since $v_{i+1}$ is accepted, we have $s_j \neq v_i$.
		By swapping $v_i$ with $v_{i+1}$, we obtain a new permutation $\pi'$ with the same reference element $s_j$.
		This is obvious if $v_i$ is not in the sampling of $\pi$. Otherwise, note that in the ordered sequences of sampling items
		from $\pi$ and $\pi'$, both $v_{i+1}$ and $v_i$ have the same position. This implies that $s_j$ is the $j$-th best sampling item in $\pi'$.
		Further, item $v_i$ is at the former position of $v_{i+1}$ in $\pi'$, thus both algorithms accept $v_i$ at this position since $v_i > v_{i+1} > s_j$.
		
		The claim follows by applying the inequality $p_i \geq p_{i+1}$ iteratively.
	\end{proof}
	
	Due to the monotonicity property, the competitive ratios of both algorithms can be easily analyzed using the following lemma.
			
		\begin{lemma}
			\label{lemma:compRatio}
			The competitive ratio of any monotone algorithm is 
			$(1/k) \sum_{i=1}^{k} p_i$.
		\end{lemma}

		\begin{proof}
			By monotonicity (\Cref{lemma:monotonicity}) and by definition of the item set, both sequences $p_1,\ldots,p_k$ and $v_1,\ldots,v_k$ are sorted decreasingly. 
			Let $\OPT = \sum_{i=1}^{k} v_i$ and
			$\E{\mathcal{A}}$ be the expected sum of the items accepted by the monotone algorithm. 
			Chebyshev's sum inequality \cite{graham1994concrete} states that if 
			$a_1 \geq a_2 \geq \ldots \geq a_n$ and $b_1 \geq b_2 \geq \ldots \geq b_n$, then
			$\sum_{i=1}^{n} a_i b_i \geq (1/n) \left( \sum_{i=1}^{n} a_i\right) \left( \sum_{i=1}^{n} b_i\right)$.
			Applying this inequality yields
			\[
			\E{\mathcal{A}} 
			= \sum_{i=1}^{n} p_i v_i 
			\geq \sum_{i=1}^{k} p_i v_i 
			\geq \frac{1}{k} \left( \sum_{i=1}^{k} v_i \right) \left( \sum_{i=1}^{k} p_i \right)
			= \left( \frac{1}{k} \sum_{i=1}^{k} p_i \right) \OPT \,.
			\]
			Note that the above inequalities are tight: Consider a set of items where the top $k$ items are basically identical, and all remaining items are close to zero. More formally, set $v_i=1 - i \varepsilon$ for $i \in \intInter{1}{k}$	
			and $v_i = i \varepsilon$ for $i \in \intInterLeftOpen{k}{n}$, where $\varepsilon \to 0$.
			Then, the competitive ratio is exactly
			$(1/k) \sum_{i=1}^{k} p_i$.
			\end{proof}
			The same argument is used in \cite{DBLP:journals/mor/BuchbinderJS14} to show that any monotone algorithm for $(k,k)$-secretary corresponds to an algorithm for $k$-secretary of the same competitive ratio.

\section{Analysis of SINGLE-REF}
\label{sec:simple}

In this section we analyze our proposed algorithm \textsc{single-ref}. 
Recall that this algorithm uses $s_r$, the $r$-th best sampling item, as the threshold for accepting items.
As implied by \Cref{lemma:compRatio}, only the $k$ largest items $v_1,\ldots,v_k$ 
contribute to the objective function. One essential idea of our approach is to separate the set of top-$k$ items into two classes according to the following definition.
\begin{definition}
	We say that item $v_i$ is \textit{dominating} if $i \leq r$, and \textit{non-dominating} if $r+1 \leq i \leq k$.
\end{definition}
The crucial property of dominating items becomes clear in the following scenario:
Assume that any dominating item $v$ occurs after the sampling phase. 
Since $s_r$ is the $r$-th best item from the sampling phase, it follows that $v > s_r$. 
That is, each dominating item outside the sampling beats the reference item. Therefore, 
there are only two situations when dominating items are rejected:  Either they appear before position $t$, 
or after $k$ accepted items.

\subsection{Dominating Items}
\label{sec:dominating}
\noindent
In \Cref{lemma:PrJthAccept} we compute the acceptance probability for dominating items.
Since the algorithm cannot distinguish any two dominating items, each dominating item has the same probability.
\begin{lemma}
	\label{lemma:PrJthAccept}
	Let $v_d$ be a dominating item and $j \in \intInterRightOpen{0}{k}$.
	We have
	\[
		p_d^{(j+1)} = \frac{\kappa \tau}{n} \sum_{i=t+j}^{n} \binom{i-t}{j} \frac{1}{\fallingfactorial{(i-1)}{r+j}} \,,
	\]
	where $\kappa = \fallingfactorial{(r-1+j)}{j}$ and $\tau = \fallingfactorial{(t-1)}{r}$.
\end{lemma}

\begin{proof}
	Let $E_j(z,i)$ be the event that \textsc{single-ref} accepts $v_d$ as $(j+1)$-th item at position
	$i=\pos(v_d)$ and $s_r$ has rank $z$ (thus $s_r=v_z$) in a random permutation.
	Note that there must be elements $s_1, \ldots, s_{r-1}$ of rank smaller than $z$ in the sampling.
	Similarly, there must be $j$ accepted elements $a_1, \ldots, a_j$ of rank smaller than $z$ outside the sampling, but before position $i$.
	
	The proof is in several steps.
	We first consider a stronger event $\tilde{E}_j(z,i)$. 
	Later, we show how the probability of $E_j(z,i)$ can be obtained from $\tilde{E}_j(z,i)$. 
	In the end, the law of total probability yields $p_d^{(j+1)}$.
	
	\subparagraph{Analysis of $\tilde{E}_j(z,i)$.}
	
	Event $\tilde{E}_j(z,i)$ is defined as $E_j(z,i)$ with additional position constraints (see \Cref{fig:eventE}):
	Elements $s_1,\ldots,s_r$ are in this order at the first $r$ positions and
	elements $a_1,\ldots,a_j$ are in this order at the $j$ positions immediately before $v_d$.
	Therefore, $\tilde{E}_j(z,i)$ holds if and only if the random input sequence satisfies the following conditions:
	\begin{enumerate}[(i)]
		\item $\pos(s_\ell)=\ell$ for  $\ell \in [r]$, $\pos(a_m) =i-j+m-1$  for $m \in [j]$, and $\pos(v_d)=i$.
		\item Elements $s_1,\ldots,s_{r-1}$ have rank smaller than $z$
		\item Elements $a_1\ldots,a_j$ have rank smaller than $z$
		\item All remaining items at positions $r+1,\ldots,i-j-1$ have rank greater than $z$.
	\end{enumerate}
	
	\begin{figure}
		\centering
		\includegraphics[width=0.7\textwidth]{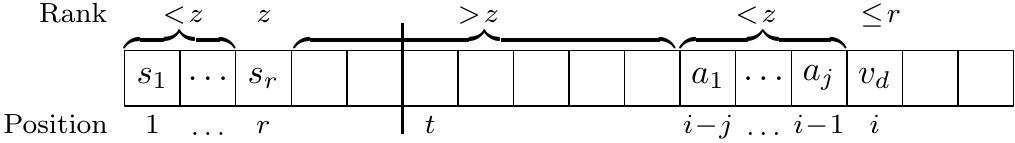}
		\caption{Event $\tilde{E}_j(z,i)$ considered in the proof of \Cref{lemma:PrJthAccept}.}
		\label{fig:eventE}
	\end{figure}

	As described in \Cref{sec:randomArrivalOrder}, we think of sequentially drawing the elements for the positions $1,\ldots,r$ first, then $i-j,\ldots, i$, and finally $r+1,\ldots,i-j-1$.
	The probability for (i) is 
	\[\beta := \prod_{\ell = 0}^{j+r} \frac{1}{n- \ell} = \frac{1}{\fallingfactorial{n}{j+r+1}} \,,\]
	since each item has the same probability to occur at each remaining position.
	In (ii), the $r-1$ elements can be chosen out of $z-2$ remaining items of rank smaller than $z$
	(since $v_d$ is dominating and was already drawn). Therefore, we get a factor of $\binom{z-2}{r-1}$.
	After this step, there remain $z-2-(r-1)=z-r-1$ elements of rank smaller than $z$, so
	we get factor $\binom{z-r-1}{j}$ for step~(iii).
	
	Finally, the probability of (iv) can be formulated using Fact~\ref{fact:BlueBalls}.
	Note that at this point, there remain $n-(1+r+j)$ items and no item of rank greater than $z$ has been drawn so far. 
	In terms of the random experiment from Fact~\ref{fact:BlueBalls}, we draw $K = i-j-r-1$ balls (items)
	from an urn of size $N=n-(1+r+j)$ where $M=n-z$ balls are blue (rank greater than $z$).
	Hence, the probability for (iv) is $H := h(n-r-j-1,n-z,i-j-r-1)$.
	Therefore, we obtain
	\begin{equation*}
	\Pr{\tilde{E}_j(z,i)} = \beta \cdot \binom{z-2}{r-1} \binom{z-r-1}{j} \cdot H \,.
	\end{equation*}
	This term can be simplified further by applying (\ref{rule:trinomialRevision}) and (\ref{rule:symmetry}).
	Let $R = z-2$, $K=r-1$, and $M=j+r-1$. It holds that
	\begin{multline*}
	\binom{z\!-\!2}{r\!-\!1} \binom{z\!-\!r\!-\!1}{j} 
	= \binom{R}{K} \binom{R\!-\!K}{M\!-\!K} 
	\overset{\text{(\ref{rule:trinomialRevision})}}{=} \binom{R}{M} \binom{M}{K} 
	\overset{\text{(\ref{rule:symmetry})}}{=} \binom{R}{M} \binom{M}{M\!-\!K} \\
	= \binom{z\!-\!2}{j\!+\!r\!-\!1} \binom{j\!+\!r\!-\!1}{j}.
	\end{multline*}
	Let $\kappa = \fallingfactorial{(j+r-1)}{j}$, then $\binom{j+r-1}{j} = \kappa / j!$ and we get
	\[\Pr{\tilde{E}_j(z,i)} = \frac{\beta \kappa}{j!} \cdot \binom{z-2}{j+r-1} \cdot H \,. \]
	
	\subparagraph*{Relating $\tilde{E}_j(z,i)$ to $E_j(z,i)$.}
	In contrast to $\tilde{E}_j(z,i)$, in the event $E_j(z,i)$, the elements 
	$s_1,\ldots,s_r$ can have any positions in $[t-1]$ and $a_1\ldots,a_j$ can have any positions in $\intInterRightOpen{t}{i}$.
	In the random order model, the probability of an event depends linearly on the number of
	permutations for which the event happens. Hence, we can multiply the probability with corresponding
	factors $\fallingfactorial{(t-1)}{r} =: \tau$ and 
	$\fallingfactorial{(i-t)}{j} = \binom{i-t}{j} j!$ and get
	\[ \Pr{E_j(z,i)} = \binom{i-t}{j} \tau j! \cdot \Pr{\tilde{E}_j(z,i)} \,. \]
	
	\subparagraph*{Relating $E_j(z,i)$ to $p_d^{(j+1)}$.}
	As the final step, we sum over all possible values for $i$ and $z$ to obtain $p_d^{(j+1)}$.
	The position $i = \pos(v_d)$ ranges between $t+j$ and $n$, while the reference rank $z$ is
	between $r+j+1$ (there are $r-1$ sampling elements and $j+1$ accepted elements of rank less than $z$) and $n$. 
	Thus we get:
	\begin{align}
	p_d^{(j+1)} 
	&= \sum_{i=t+j}^{n} \sum_{z=r+j+1}^{n} \Pr{E_j(z,i)} \nonumber \\
	&= \tau j! \sum_{i=t+j}^{n} \binom{i-t}{j} \sum_{z=r+j+1}^{n}  \Pr{\tilde{E}_j(z,i)}   \nonumber \\
	&= \beta \kappa \tau \sum_{i=t+j}^{n} \binom{i-t}{j} \sum_{z=r+j+1}^{n}   
	\binom{z-2}{j+r-1} \cdot H  \nonumber \\
	&= \beta \kappa \tau \sum_{i=t+j}^{n} \binom{i-t}{j} 
	\frac{1}{\binom{n-r-j-1}{i-j-r-1}}
	\sum_{z=r+j+1}^{n}   
	\binom{z-2}{j+r-1} \binom{n-z}{i-j-r-1}	\label{eq:PrEjFirst} \,.
	\end{align}
	The sum over $z$ in (\ref{eq:PrEjFirst}) can be resolved using (\ref{rule:sumOfProducts}).
	Let $L = n-r-j-1$, $N=W=r+j-1$, and $M=i-j-r-1$. 
	In order to apply (\ref{rule:sumOfProducts}) we need to verify $L,M \geq 0$ and $N \geq W \geq 0$.
	We can assume $k \leq n/2$, since for $k > n/2$, there exist a trivial $(1/2)$-competitive algorithm.
	Therefore, we have $L = n-r-j-1 \geq n-k-(k-1)-1 = n-2k \geq 0$. 
	Further, $i \geq t+j$, thus $i-j \geq t \geq k + 1 \geq r+1$ which implies $M \geq 0$.
	The condition $N \geq W \geq 0$ holds trivially. By (\ref{rule:sumOfProducts}) we obtain
	\begin{align}
	&\sum_{z=r+j+1}^{n}   	\binom{z-2}{j+r-1} \binom{n-z}{i-j-r-1}  \nonumber \\
	&= \sum_{z=0}^{n-r-j-1}   	\binom{r+j-1+z}{j+r-1} \binom{n-r-j-1-z}{i-j-r-1}  \nonumber \\ 
	&= \sum_{z=0}^{L}   	\binom{W+z}{N} \binom{L-z}{M} 
	= \binom{L+W+1}{M+N+1} 
	= \binom{n-1}{i-1} \,.
	\label{eq:sumOfProductsn-1i-1}
	\end{align}
	
	By inserting (\ref{eq:sumOfProductsn-1i-1}) into (\ref{eq:PrEjFirst}), we obtain the quotient of binomial coefficients 
	$\binom{n-1}{i-1} / \binom{n-r-j-1}{i-j-r-1}$, which can be simplified further using (\ref{rule:trinomialRevision}) since
	\[
	\binom{n-1}{i-1} \Big/ \binom{n-1-(r+j)}{i-1-(r+j)}	
	= \binom{n-1}{r+j} \Big/ \binom{i-1}{r+j} 
	= \frac{\fallingfactorial{(n-1)}{r+j}}{\fallingfactorial{(i-1)}{r+j}} \,.
	\]
	Recall $\beta = 1 / \fallingfactorial{n}{j+r+1}$, thus
	$\fallingfactorial{(n-1)}{r+j} \cdot \beta = 1/n$.
	Together with (\ref{eq:PrEjFirst}) we get
	\begin{align}
	\label{eq:prEjexact}
	p_d^{(j+1)}
	&= \beta \kappa \tau \cdot \fallingfactorial{(n-1)}{r+j}
	\sum_{i=t+j}^{n} \binom{i-t}{j} \frac{1}{\fallingfactorial{(i-1)}{r+j}} \nonumber \\
	&= \frac{\kappa \tau}{n} 
	\sum_{i=t+j}^{n} \binom{i-t}{j} \frac{1}{\fallingfactorial{(i-1)}{r+j}} \,,
	\end{align}
	which concludes the proof.
\end{proof}

\subsection{Non-Dominating Items}
\label{sec:non-dominating}

It remains to consider the acceptance probabilities of the non-dominating items $v_{r+1},\ldots,\allowbreak v_k$.
Interestingly, these probabilities can be related finally to those for dominating items.
First, we obtain the following result.

\begin{lemma}
	\label{lemma:ProbNonDomHoricontal}
	Let $v_{r+i}$ be a non-dominating item with $i \in [k-r]$ and let $j \in [i]$.
	We have  $p_{r+i}^{(j)} = p_{r+i}^{(i+1)}$.
	\end{lemma}
	\begin{proof}
		We construct a bijective function $f \colon P \to Q$ where $P$ (resp. $Q$) is the set of permutations where $v_{r+i}$ is the $j$-th (resp. $(i+1)$-th) accept.
		
		Let $\pi \in P$. First, we argue that the algorithm accepts at least $i+1$ elements in $\pi$.
		As $v_{r+i}$ is accepted, $s_r < v_{r+i}$ and thus all elements from the set $\mathcal{S} = \{v_1,\ldots,v_{r+i}\}$ beat $s_r$.
		Since $s_r$ is the $r$-th best element in the sampling, at most $r-1$ elements from $\mathcal{S}$ can
		be part	of the sampling. Consequently, at least $r+i - (r-1) = i+1$ elements from $\mathcal{S}$, including $v_{r+i}$, are accepted.
		
		Now, let $a_1,\ldots,a_{i+1}$ denote the first $i+1$ accepts, where $a_j = v_{r+i}$.
		The function $f$ can be defined as follows: Swap the positions of $a_1,\ldots,a_{i+1}$ in a cyclic shift, such that
		$a_j = v_{r+i}$ is at the former position of $a_{i+1}$. This yields a permutation $f(\pi)$ where $v_{r+i}$ is the $(i+1)$-th accept.
		Note that $f$ is bijective as the cyclic shift can be reversed.
		\end{proof}

		By the following lemma, the remaining probabilities can be related to corresponding probabilities for dominating items.
		\begin{lemma}
			\label{lemma:ProbNonDomVertical}
			Let $v_{r+i}$ be a non-dominating item with $i \in [k-r]$ and let $j \in [k-i]$. 
			Let  $v_d$ be any dominating item. It holds that
			$p_{r+i}^{(i+j)} = p_d^{(i+j)}$.
			\end{lemma}
			\begin{proof}
				Again, we prove the claim by defining a bijective mapping $f \colon P \to Q$ where
				$P$ is the set of permutations where $v_{r+i}$ is the $(i+j)$-th accept and
				$Q$ is the set of permutations where  $v_d$ is the $(i+j)$-th accept.
				For any $\pi \in P$, let $f(\pi)$ be obtained from $\pi$ by swapping $v_{r+i}$ with $v_d$.
				
				We first show that $f \colon P \to Q$. 
				To that end, consider any fixed $\pi \in P$. As $v_{r+i}$ is accepted, $s_r < v_{r+i}$.
				We can argue that $s_r$ is the $r$-th best sampling element in $f(\pi)$ as well:
				This holds clearly if no item is moved out of or into the sampling by $f$.
				Otherwise, $f$ moves $v_{r+i}$ into the sampling and $v_d$ outwards.
				But since $s_r < v_{r+i} < v_d$, this does not change the role of $s_r$ as the $r$-th best sampling element.
				Therefore, $\pi$ and $f(\pi)$ have the same reference element $s_r$. Further, $v_d$ is accepted in $f(\pi)$ at the former position of $v_{r+i}$. Finally, we observe that $f$ is bijective, since $v_{r+i}$ and $v_{d}$ have unique ranks.
				\end{proof}

\subsection{Competitive Ratio}
\label{sec:SRcompRatio}
The following main theorem states the exact competitive ratio of \textsc{single-ref}. By the results from \Cref{sec:dominating,sec:non-dominating}, the final term only depends on the acceptance probabilities of dominating items.

\begin{theorem}
\label{theo:SRcompRatio}
	The competitive ratio of \textsc{single-ref} is
	\[
	\frac{1}{k} \sum_{j=1}^{k} \gamma_j \cdot p_1^{(j)}
	\hspace*{15pt}
	\text{where }
	\gamma_j =
	\begin{cases}
	r+2\cdot(j-1) & \text{if } j \leq k-r+1 \\
	k & \text{else.} \\	
	\end{cases}
	\]
	
\end{theorem}
\begin{proof}
	According to \Cref{lemma:compRatio}, the competitive ratio is given by
	\begin{equation}
	\label{eq:SRcompRatio}
	\frac{1}{k} \sum_{i=1}^{k} p_i = 
	\frac{1}{k} \left( \sum_{i=1}^{r} p_i + \sum_{i=1}^{k-r} p_{r+i} \right) \,,
	\end{equation}
	where we split the sum according to dominating and non-dominating items.
	By \Cref{lemma:PrJthAccept}, $p_i^{(j)}=p_1^{(j)}$ holds for any dominating item $v_i$ and $j \in [k]$.
	Recall that $p_i = \sum_{j=1}^{k} p_i^{(j)}$ for all $i \in [k]$. Therefore,
	\begin{equation}
	\label{eq:SRcompRatioDom}
	\sum_{i=1}^{r} p_i = r \sum_{j=1}^{k} p_i^{(j)} = \sum_{j=1}^{k} r p_1^{(j)} \,.
	\end{equation}
	Now, we consider the sum for non-dominating items.	We have 
	\begin{equation}
	\label{eq:SRcompRatioNonDom1}
	\sum_{i=1}^{k-r} p_{r+i} 
	= \sum_{i=1}^{k-r} \sum_{j=1}^{k} p_{r+i}^{(j)} 
	= \sum_{i=1}^{k-r} \sum_{j=1}^{i} p_{r+i}^{(j)} + \sum_{i=1}^{k-r} \sum_{j=1}^{k-i} p_{r+i}^{(i+j)} \,.
	\end{equation}
	Next, we simplify the second last sum of \Cref{eq:SRcompRatioNonDom1}.
	From \Cref{lemma:ProbNonDomHoricontal,lemma:ProbNonDomVertical} it follows that 
	\begin{equation}
	\label{eq:SRcmopRatioNonDom2}
	\sum_{i=1}^{k-r} \sum_{j=1}^{i} p_{r+i}^{(j)} 
	= \sum_{i=1}^{k-r} i p_{r+i}^{(i+1)} 
	= \sum_{i=1}^{k-r} i p_1^{(i+1)} 
	= \sum_{j=1}^{k-r+1} (j-1)  p_1^{(j)} \,.		
	\end{equation}
	The last sum in \Cref{eq:SRcompRatioNonDom1} can be simplified using \Cref{lemma:ProbNonDomVertical} followed by algebraic manipulations.
	\begin{multline}
	\label{eq:SRcmopRatioNonDom3}
	\sum_{i=1}^{k-r} \sum_{j=1}^{k-i} p_{r+i}^{(i+j)}
	= \sum_{i=1}^{k-r} \sum_{j=1}^{k-i} p_{1}^{(i+j)}
	= \sum_{i=1}^{k-r} \sum_{j=i+1}^{k} p_{1}^{(j)}
	= \sum_{j=1}^{k-r+1} (j-1) p_1^{(j)} + \sum_{j=k-r+2}^{k}  (k-r) p_1^{(j)} \,.
	\end{multline}
	Combining \Cref{eq:SRcompRatio,eq:SRcompRatioDom,eq:SRcompRatioNonDom1,eq:SRcmopRatioNonDom2,eq:SRcmopRatioNonDom3}
	yields the claim.
\end{proof}

\subsection{Dominating Items -- Asymptotic Setting}
As we have seen in \Cref{theo:SRcompRatio}, the competitive ratio mainly depends on the probabilities $p_1^{(j)}$.
However, the term from Equation~(\ref{eq:prEjexact}) is cumbersome and hard to optimize over $r$ and $t$.
The goal of this subsection is to derive a lower bound for $p_1^{(j)}$ assuming that $n$ is large enough.
For this purpose, we assume $t-1 = cn$ for $c \in (0,1)$, i.e., the sampling length is some constant $c$ of the input length.
Further, we assume that $k \in o(n)$.
We obtain the following lemma.

\begin{lemma}
	\label{lemma:dominatingItemsAsymptotic}	
	Let $j \in \intInterRightOpen{0}{k}$.
	For $\ell \in \intInter{0}{j}$, define
	$\beta_\ell := (-1)^\ell \binom{j}{\ell}$ and $\alpha_\ell := \binom{j+r-1}{\ell+r-1}$.
	Assuming $t-1 = cn$, it holds that
	\[
	p_1^{(j+1)} \geq \begin{cases}
	c  \cdot \left(\ln \frac{1}{c} - \sum_{\ell=1}^{j} \beta_\ell \frac{c^\ell -1}{\ell}\right) - o(1)
	& \text{if $r=1$} \\
	\frac{c}{r-1} \cdot \left( 1 - c^{r-1} \cdot \sum_{\ell=0}^{j} \alpha_\ell  (1-c)^{j-\ell} c^\ell \right) - o(1)
	& \text{if $r \geq 2$.} \\	
	\end{cases}
	\]
\end{lemma}

\begin{proof}[Proof of \Cref{lemma:dominatingItemsAsymptotic}]
	Let $S:= \sum_{i=t+j}^{n} \binom{i-t}{j} \frac{1}{\fallingfactorial{(i-1)}{r+j}}$ be the sum from Equation~(\ref{eq:prEjexact}). 
	First, we obtain a lower bound for $S$ using $(n-k)^k < \fallingfactorial{n}{k} < n^k$ and similar inequalities:
	\begin{multline}
	\label{eq:singleRefS}
	S 
	= \sum_{i=t+j}^{n} \binom{i-t}{j} \frac{1}{\fallingfactorial{(i-1)}{r+j}}
	= \frac{1}{j!} \sum_{i=t+j}^{n} \frac{\fallingfactorial{(i-t)}{j}}{\fallingfactorial{(i-1)}{r+j}}
	> \frac{1}{j!} \sum_{i=t+j}^{n} \frac{(i-t-j+1)^j}{(i-1)^{r+j}} \\
	= \frac{1}{j!} \sum_{i=1}^{n-t-j+1} \frac{i^j}{(i+t+j-2)^{r+j}}
	= \frac{1}{j!} \sum_{i=1}^{m} f(i) \,,
	\end{multline}
	where we defined $f(i) := i^j / (i+y)^{r+j}$ with $y:=t+j-2$ and $m:=n-t-j+1$.
	Unfortunately, $f$ is not necessarily monotone, hence we cannot apply Fact~\ref{lemma:approxSumByIntegral} directly to bound $S$ by a corresponding integral. However, $f$ has a single maximum point $z = \frac{jy}{r}$ and is monotone increasing (resp.\ monotone decreasing) for $i \leq z$ (resp.\ $i \geq z$). We prove this property in \Cref{lemma:fMonotoneSingleMaximum} in \ref{app:single-ref}.
	This allows to split $S$ into two monotone parts.
	\begin{align}
	\sum_{i=1}^{m} f(i) 
	&= 	\sum_{i=1}^{\lfloor z \rfloor} f(i) + \sum_{i= \lfloor z \rfloor +1}^{m} f(i) \nonumber \\
	&\geq \int_0^{\lfloor z \rfloor} f(i) \intD{i} + \int_{\lfloor z \rfloor +1}^{m+1} f(i) \intD{i} \nonumber & \text{Fact~\ref{lemma:approxSumByIntegral}, \Cref{lemma:fMonotoneSingleMaximum}B} \\
	&= \int_0^{m+1} f(i) \intD{i} - \int_{\lfloor z \rfloor}^{\lfloor z \rfloor +1} f(i) \intD{i} \nonumber \\
	&= \int_0^{m+1} f(i) \intD{i} - f(z) & \text{\Cref{lemma:fMonotoneSingleMaximum}A}\,.
	\label{eq:sumAsIntegrals}
	\end{align}
	Therefore, if $F$ is a function such that $\int_0^{m+1} f(i) \intD{i} = F(m+1) - F(0)$, we obtain from \Cref{lemma:PrJthAccept}, Inequality (\ref{eq:sumAsIntegrals}), and the previous observation
	\begin{equation}
		\label{eq:sumAsAntiDerivativesAbstract}
		p_1^{(j+1)}
		\geq \frac{\kappa \tau}{n} S 
		> \frac{\kappa \tau }{n j!} \left( F(m+1) - F(0) \right)  - \frac{\kappa \tau }{n j!} f(z) \,.
	\end{equation}	
	We first argue that the term $(\kappa \tau) / (n j!) \cdot f(z)$ from Inequality~(\ref{eq:sumAsAntiDerivativesAbstract}) vanishes for $n \to \infty$.
	It holds that $\kappa \tau = \fallingfactorial{(r+j-1)}{j} \cdot \fallingfactorial{(t-1)}{r} < (r+j)^j \cdot (t-1)^r$. Moreover,
	\[
	f(z) 
	= \frac{z^j}{(z+y)^{r+j}} 
	< \frac{z^j}{y^{r+j}} 
	= \frac{\left( \frac{jy}{r}\right)^j}{y^{r+j}} 
	= \left( \frac{j}{r} \right)^j \cdot \frac{1}{y^r}
	< \frac{j^j}{(t-1)^r \cdot (1 - \frac{1}{t-1} )^r} \,,
	\]
	where the last inequality follows from $r \geq 1$ and $y^r \geq (t-2)^r = (t-1)^r \cdot (1 - \frac{1}{t-1})^r$.
	Therefore,
	\begin{multline}
	\label{eq:o1}
	\frac{\kappa \tau }{n j!} f(z)
	< \frac{(r+j)^j \cdot (t-1)^r}{n j!} \cdot \frac{j^j}{(t-1)^r \cdot (1 - \frac{1}{t-1} )^r}
	= \frac{{(rj+j^2)}^j}{j!} \cdot \frac{1}{n \cdot (1 - \frac{1}{cn} )^r} \\
	= o(1) \,.
	\end{multline}

	In the remainder of the proof, we consider the cases $r=1$ and $r \geq 2$ separately.
	For both cases we use an appropriate antiderivative $F$, where we prove $F'(i)=f(i)$ in \Cref{lemma:fAntiderivativeR1,lemma:fAntiderivativeR2} in \ref{app:single-ref}.

	\paragraph{Case $r=1$}
		We observe first that in \Cref{eq:sumAsAntiDerivativesAbstract}, the factor $\frac{\kappa \tau}{n j!}$ resolves to $c$ as
		\[
		\frac{\kappa \cdot \tau}{n j!}
		= \frac{\fallingfactorial{(j+r-1)}{j} \cdot \fallingfactorial{(t-1)}{r}}{n j!}
		= \frac{\fallingfactorial{j}{j} \cdot \fallingfactorial{(t-1)}{1}}{n j!}
		= \frac{j! \cdot (t-1)}{n j!}
		= c \,.
		\]
		With $\beta_\ell = (-1)^\ell \binom{j}{\ell}$ for $1 \leq \ell \leq j$ and
		\[
		F(i) = \ln(i+y) - \sum_{\ell=1}^{j} \frac{\beta_\ell}{\ell} \cdot \left( \frac{y}{i+y}\right)^\ell \,,
		\]		
	it holds that
	\begin{align*}
	&\qquad F(m+1) - F(0) \\
	&= \left( \ln(m+1+y) - \sum_{\ell=1}^{j} \frac{\beta_\ell}{\ell} \cdot \left( \frac{y}{m+1+y} \right) ^\ell \right) 
	- \left( \ln y - \sum_{\ell=1}^{j} \frac{\beta_\ell}{\ell} \right) \\
	&= \ln \left(  \frac{m+1+y}{y}\right) - \sum_{\ell=1}^{j} \frac{\beta_\ell}{\ell} \cdot  \left( \left(\frac{y}{m+1+y}\right)^\ell - 1 \right) \\
	&= \ln \left(  \frac{n}{y}\right) - \sum_{\ell=1}^{j} \frac{\beta_\ell}{\ell} \cdot  \left( \left(\frac{y}{n}\right)^\ell - 1 \right) \\
	&= \ln \frac{1}{c} - \left( \sum_{\ell=1}^{j} \frac{\beta_\ell}{\ell} \cdot  \left( c^\ell - 1 \right) \right) - o(1) \,,
	\end{align*}
	where we used $m+1+y = n$ and $y=t-j-2 = cn -j - 1$.
	Combining with Inequality (\ref{eq:sumAsAntiDerivativesAbstract}) yields the claim for $r=1$.

	\paragraph{Case $r \geq 2$}
	Define $\alpha_\ell = \binom{j+r-1}{\ell+r-1}$ for $0 \leq \ell \leq j$ and let
	\[
	F(i) = - \frac{\sum_{\ell=0}^{j} \alpha_\ell  i^{j-\ell} y^\ell}{\alpha_0 (r-1) (i+y)^{r+j-1}} \,.
	\]
	Further, let $G(i) = - \alpha_0 \cdot (r-1) \cdot F(i)$.
	Since $\alpha_0 = \binom{j+r-1}{r-1} = \kappa / j!$ and $\tau / n = \fallingfactorial{(t-1)}{r} / n = c \cdot \fallingfactorial{(t-2)}{r-1}$, we have
	\begin{equation}
	\label{eq:SRasymR2-1}
	\frac{\kappa \tau }{n j!} \left( F(m+1) - F(0) \right)
	%&= \frac{\kappa \tau }{n j!} \left( \frac{G(0)}{\alpha_0 (r-1)} - \frac{G(m+1)}{\alpha_0 (r-1) } \right) \nonumber \\
	= \frac{c}{r-1} \cdot \fallingfactorial{(t-2)}{r-1} \cdot \left(G(0) - G(m+1) \right)	\,.
	\end{equation}
	Now, it holds that
	\begin{align}
	\label{eq:SRasymR2-2}
	G(0) 
	= \frac{\sum_{\ell=0}^{j} \alpha_\ell  \cdot 0^{j-\ell} \cdot y^\ell}{y^{r+j-1}}
	= \frac{\alpha_j \cdot y^j}{y^{r+j-1}}
	= \frac{1}{y^{r-1}}
	= \frac{1}{{(t+j-2)}^{r-1}}
	\end{align}
	and further
	\begin{align}
	\label{eq:G0}
	\fallingfactorial{(t-2)}{r-1} \cdot G(0)
	&\geq \frac{{(t-r)}^{r-1}}{{(t+j-2)}^{r-1}}
	= \left( 1 - \frac{j+r-2}{t+j-2} \right)^{r-1}
	= 1 - o(1) \,.
	\end{align}
	
	Finally, we analyze the term $\fallingfactorial{(t-2)}{r-1} \cdot G(m+1)$.	
	Using $m=n-t-j+1$ and $y = t+j-2 = cn + j-1$, we obtain
	\begin{align}
	\label{eq:Gm1}
	\fallingfactorial{(t-2)}{r-1} \cdot G(m+1) 
	&= \fallingfactorial{(t-2)}{r-1} \cdot \frac{\sum_{\ell=0}^{j} \alpha_\ell  \cdot (m+1)^{j-\ell} \cdot y^\ell}{(m+1+y)^{r+j-1}} \nonumber \\
	&\leq \fallingfactorial{(t-2)}{r-1} \cdot \frac{\sum_{\ell=0}^{j} \alpha_\ell  \cdot (n \cdot (1-c) + 1)^{j-\ell} \cdot (cn + j-1)^\ell}{n^{r+j-1}} \nonumber \\
	&\leq \fallingfactorial{(t-2)}{r-1} \cdot \frac{\left( \sum_{\ell=0}^{j} \alpha_\ell  \cdot (n \cdot (1-c))^{j-\ell} \cdot (cn)^\ell \right)}{n^{r+j-1} } + o(1) \nonumber \\	
	&< (t-1)^{r-1} \cdot \frac{\left( \sum_{\ell=0}^{j} \alpha_\ell  \cdot n^{j-\ell} \cdot (1-c)^{j-\ell} \cdot c^\ell n^\ell \right)}{n^{r+j-1} } + o(1)\nonumber \\		
	&= c^{r-1} \cdot \left(\sum_{\ell=0}^{j} \alpha_\ell  \cdot (1-c)^{j-\ell} \cdot c^\ell\right) + o(1) \,.
	\end{align}
	Combining \Cref{eq:sumAsAntiDerivativesAbstract,eq:o1,eq:G0,eq:Gm1,eq:SRasymR2-1} concludes the proof.
\end{proof}

\subsection{Competitive Ratio -- Asymptotic Setting}

	\begin{figure}[t]
		\centering
		%\resizebox{0.8\linewidth}{!}{\input{plot.tex}}
		\includegraphics{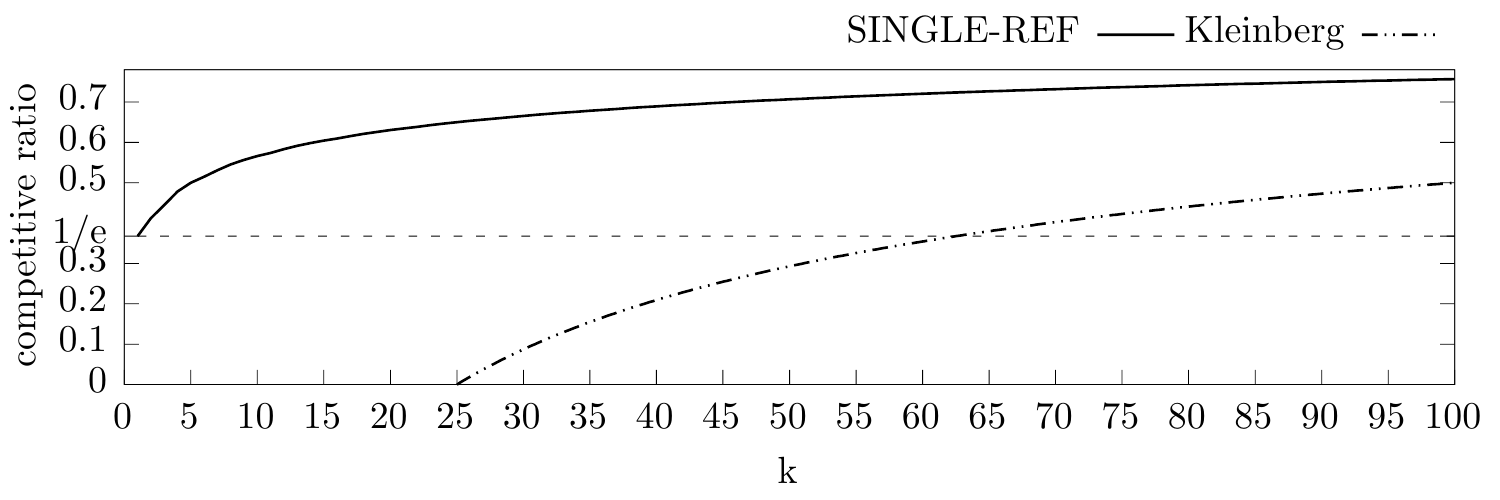}
		\caption{Comparison of our algorithm \textsc{single-ref} and the algorithm by Kleinberg 
			\cite{DBLP:conf/soda/Kleinberg05}.}
		\label{fig:singlerefCompetitiveRatiosFull}
	\end{figure}

	The competitive ratio of \textsc{single-ref} in the asymptotic setting $n \to \infty$ can be evaluated using \Cref{lemma:dominatingItemsAsymptotic} in combination with \Cref{theo:SRcompRatio}. For $k \in \intInter{1}{100}$, we optimized the resulting objective function over $r$ and $c$	numerically. 
	As shown in \Cref{fig:singlerefCompetitiveRatiosFull},  \textsc{single-ref} reaches competitive ratios of up to $0.75$ and outperforms the algorithm by Kleinberg \cite{DBLP:conf/soda/Kleinberg05} on this interval.
	The optimal parameters for $k \in \intInter{1}{100}$ and the resulting competitive ratios can be found in \Cref{tab:comeptitiveRatios20Full} of \ref{app:single-refTable}.

\section{OPTIMISTIC for \texorpdfstring{$k=2$}{k=2}}
\label{sec:optimistic}

Let $\mathcal{A}_2$ denote the \textsc{optimistic} algorithm with $k=2$ and parameters $r$, $t$ in the following.
To analyze the relevant probabilities $p_1$ and $p_2$, we again relate these probabilities to corresponding sets (see \Cref{sec:randomArrivalOrder}).
For $i \in \{1,2\}$, let $P_i$ be the set of permutations in which $\mathcal{A}_2$ accepts $v_i$.

\subsection{Acceptance Probability of \texorpdfstring{$v_2$}{v2}}

In \Cref{lemma:optimisticP2}, we show a surprising relation between \textsc{optimistic} for $k=2$ and \textsc{single-ref} for $k=r=1$:
Assuming that both algorithms have the same sampling length of $t-1$, the probability that \textsc{optimistic} accepts $v_2$ is exactly the probability that \textsc{single-ref} accepts $v_1$.
The proof uses a sophistically tailored bijection between two respective sets of permutations.

\begin{lemma}
	\label{lemma:optimisticP2}
	Let $\mathcal{A}_1$ be the \textsc{single-ref} algorithm with parameters $k=r=1$ and $t$.
	It holds that
	$\Pr{\mathcal{A}_2 \text{ accepts } v_2} = \Pr{\mathcal{A}_1 \text{ accepts } v_1}$.
\end{lemma}
\begin{proof}
	Let $\Omega$ be the set of all $n!$ permutations and $Q_1 \subset \Omega $ be the set of permutations where $\mathcal{A}_1$ accepts $v_1$.
	We prove the claim by constructing a bijective function 
	$f \colon \Omega \setminus P_2 \to \Omega \setminus Q_1$.
	
	We first investigate the two complementary sets $\Omega \setminus P_2$ and  $\Omega \setminus Q_1$.
	The set $\Omega \setminus P_2$ contains the permutations where $v_2$ is not accepted by $\mathcal{A}_2$. This occurs in exactly one of three cases: 
	$v_2$ is in the sampling, 
	$v_2$ comes behind the first accept and $v_1$ is in the sampling (then $\mathcal{A}_2$ will reject all following elements),
	or $v_2$ comes behind two elements accepted by $\mathcal{A}_2$.
	Similarly, $\Omega \setminus Q_1$ contains the permutations where $v_1$ is not accepted by $\mathcal{A}_1$. In these permutations, $v_1$ is either in the sampling, or behind
	the first accepted element.
	
	Now, fix any $\pi \in \Omega \setminus P_2$. To define $f(\pi)$, we distinguish five cases (A)-(E) based on the position of $v_2$ in $\pi$ and the accepted elements of $\mathcal{A}_2$ on input $\pi$.
	
	\begin{enumerate}[(A)]
		
		\item $v_2$ is in the sampling.
		We obtain $f(\pi)$ from $\pi$ by swapping $v_1$ with $v_2$. In $f(\pi)$, item $v_1$ is in the sampling and cannot be accepted by $\mathcal{A}_1$.
		
		\item $v_2$ comes behind the first accept and $v_1$ is in the sampling.
		Then, we have an accepted element $a_1$ with $s_2 < a_1 < v_2$.
		To obtain $f(\pi)$, we swap $v_1$ with $a_1$ and afterwards $v_1$ with $v_2$:
		\begin{center}
			\includegraphics[width=0.7\textwidth]{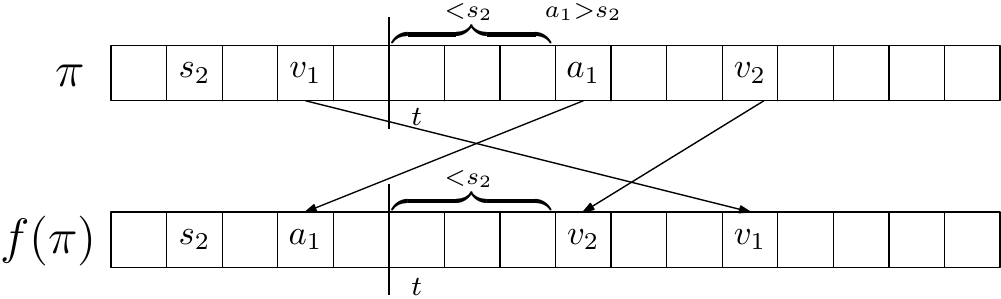}
		\end{center}
		
		Since $a_1 > s_2$, item $a_1$ is the best element in the sampling of $f(\pi)$. Particularly, the second best element $s_2$ of the sampling is maintained in this case.
		This fact will be important later. 
		Since $v_2$ is in front of $v_1$ in $f(\pi)$, algorithm $\mathcal{A}_1$ accepts $v_2$ in $f(\pi)$.
		
		\item $v_2$ comes behind two accepts and $v_1$ is the first accept.
		Then, there must be another accept $a_2$ between $v_1$ and $v_2$.
		We obtain $f(\pi)$ by swapping $v_1$ with $a_2$. Since $a_2 > s_1$, algorithm
		$\mathcal{A}_1$ accepts $a_2$ in $f(\pi)$.
		\begin{center}
			\includegraphics[width=0.7\textwidth]{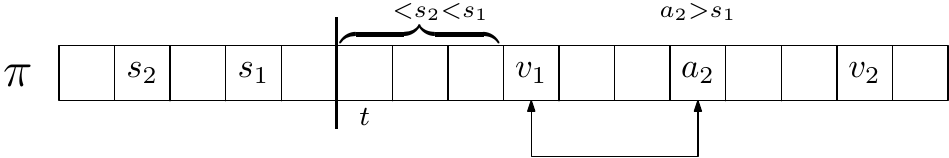}
		\end{center}		
		
		\item $v_2$ comes behind two accepts and $v_1$ is the second accept.
		Here, we define $f(\pi)$ such that $v_1$ is swapped with $v_2$.
		$\mathcal{A}_1$ accepts	$v_2$ in $f(\pi)$.
		\begin{center}
			\includegraphics[width=0.7\textwidth]{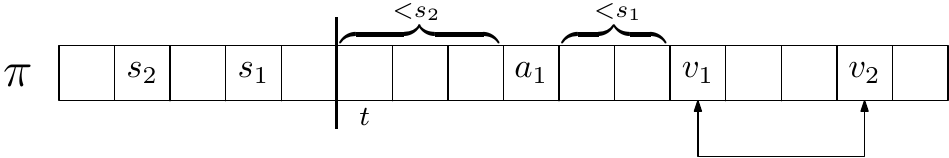}
		\end{center}	
		
		\item $v_2$ comes behind two accepts and $v_1$ is not accepted. 
		Set $f(\pi)=\pi$ in this case.
		Since there is at least one item better than $s_1$ before $v_1$ in $f(\pi)$, 
		algorithm $\mathcal{A}_1$ cannot select $v_1$ in $f(\pi)$.
	\end{enumerate}
	
	In order to show the bijectivity of $f$, we have to argue carefully following the definition of $f$ in the different cases.
	Let $\mathcal{C} = \{ A,B,C,D,E\}$ represent the set of cases (A)-(E).	
	
	\subparagraph{$f$ is injective.}
	Let $\pi_1, \pi_2 \in \Omega \setminus P_2$ with $\pi_1 \neq \pi_2$.
	We have to show $f(\pi_1) \neq f(\pi_2)$.
	Let $X,Y \in \mathcal{C}$ be such that $\pi_1$ and $\pi_2$ satisfy the conditions of case $X$ and $Y$, respectively.
	
	First, we consider the situation where $\pi_1$ and $\pi_2$ are mapped by $f$ according to the same case $X=Y$.
	In all cases, the operation defined by $f$ involves $v_1$, $v_2$, and possibly a further accepted element.
	All of these elements can be retrieved given any permutation $f(\pi)$: Items $v_1$ and $v_2$ have a unique rank,
	item $a_1$ from case $B$ is the maximum element in the sampling,
	and item $a_2$ from case $C$ is the first item after the sampling better than $s_1$.
	Hence, $f(\pi_1) = f(\pi_2)$ implies $\pi_1 = \pi_2$, which is equivalent to the claim we wanted to show.
	
	It remains to consider all pairs of cases where $X \neq Y$.
	
	\begin{description}
		\item[$X=A$, $Y \in \{B,C,D,E\}$.]
		Since item $v_1$ is moved into the sampling in case (A) and to some position after the sampling in all remaining cases (B)-(E), we immediately get $f(\pi_1) \neq f(\pi_2)$.
		
		\item[$X=E$, $Y \in \{C,D\}$.]
		Assume $f(\pi_1) = f(\pi_2)$ for contradiction. In all cases (C)-(E), the function $f$ maintains the items 
		in the sampling	phase. Particularly, $\pi_1$ and $\pi_2$ must have the same two best elements 	$s_1$, $s_2$.
		The construction in case (E) ensures that $\mathcal{A}_2$ would accept two elements in $f(\pi_1)$ before 
		position $\min \{ \pos_{f(\pi_1)} (v_1), \pos_{f(\pi_1)} (v_2)\}$, while there is only one such accept in $f(\pi_2)$.
		This contradicts $f(\pi_1) = f(\pi_2)$.
		
		\item[$X=B$, $Y \in \{D,E\}$.]
		Assume $f(\pi_1) = f(\pi_2)$ for contradiction.
		In all three cases (B),(D),(E), the function $f$ maintains the second best sampling item $s_2$.
		Similar to the previous case, we get a contradiction as there is no element before $v_2$
		better than $s_2$ in $f(\pi_1)$, while $f(\pi_2)$ has at least one such item.
		Thus, $f(\pi_1) \neq f(\pi_2)$.
		
		\item[$X=C$, $Y \in \{B,D\}$.]
		By construction, item $v_1$ is before $v_2$ in $f(\pi_1)$, while the relative order of $v_1$ and $v_2$ in $f(\pi_2)$ is the other way round.	Therefore, $f(\pi_1) \neq f(\pi_2)$.
	\end{description}
	
	\subparagraph{$f$ is surjective.}
	Let $\pi' \in \Omega \setminus Q_1$. We show that there is $\pi \in \Omega \setminus P_2$ 
	with $f(\pi)=\pi'$ .
	
	The obvious case is when
	$v_1$ is in the sampling of $\pi'$, then $\pi$ can be obtained from case (A).
	If $v_1$ is not in the sampling of $\pi'$, there must be an element before $v_1$ that $\mathcal{A}_1$ accepts.
	Consequently, $\mathcal{A}_2$ would accept at least one element before $v_1$ in $\pi'$ as well.
	Furthermore, it follows that $v_2$ is not in the sampling either, as otherwise $v_1$ would be accepted by $\mathcal{A}_1$.
	Therefore, considering $\mathcal{A}_2$ on input $\pi'$, item $v_2$ can be the first or second accept, or can follow two accepts.
	
	If $v_2$ is the first accept of $\mathcal{A}_2$, the desired $\pi$ can be constructed according
	to case (B).
	Similarly, the situation where $v_2$ is the second accept of $\mathcal{A}_2$ corresponds to case (D).
	Finally, consider the case where $v_2$ follows two accepted elements.
	If $v_1$ is the second accept, $\pi'$ was obtained from case (C), and if $v_1$ is not accepted from case (E).
	Note that $v_1$ cannot be the first accept of $\mathcal{A}_2$, as otherwise $\mathcal{A}_1$ would also accept
	$v_1$ which would contradict $\pi' \in \Omega \setminus Q_1$.
\end{proof}

\subsection{Acceptance Probability of \texorpdfstring{$v_1$}{v1}}
\label{sec:optimisticP1}

In this part, we prove $p_1 = p_2 + \delta$ for $\delta > 0$.
First, we observe that $P_2$ can be related to a set $P_1' \subset P_1$ of equal cardinality.
\begin{lemma}
	\label{lemma:P2SubsetP1}
	Let $P_1' = \{ \pi \in P_1 \mid \pos_\pi(v_2) < t \Rightarrow \mathcal{A}_2 \text{ accepts $v_1$ as the first item} \}$.
	It holds that $\card{P_1'} = \card{P_2}$.
\end{lemma}
\begin{proof}
	Let $f$ be the function that swaps $v_1$ with $v_2$ in a given input sequence.
	We first prove that $f(\pi) \in P_1'$ holds for each $\pi \in P_2$ and thus $f \colon P_2 \rightarrow P_1'$.
	
	Let $\pi \in P_2$. In $f(\pi)$, algorithm $\mathcal{A}_2$ accepts $v_1$ at position $\pos_{f(\pi)}(v_1) = \pos_\pi(v_2)$, 
	as $v_1 > v_2$. So far we showed $f(\pi) \in P_1$.
	If $\pos_{f(\pi)}(v_2) \geq t$, there is nothing to show. 
	Assuming that $\pos_{f(\pi)}(v_2) < t$, it follows that $v_1$ is the best element
	in the sampling of $\pi$. Since no item (particularly not $v_2$) beats $v_1$, but $v_2$ is accepted by
	$\mathcal{A}_2$ in $\pi$, we get that $v_2$ is the first accept in $\pi$. Hence, $v_1$ is the first accept in $f(\pi)$ and therefore $f(\pi) \in P_1'$.
	
	Clearly, $f$ is injective. To prove surjectivity, let $\pi' \in P_1'$ and let $\pi$ the permutation obtained from $\pi'$ by
	swapping (back) $v_1$ with $v_2$. If $\pos_{\pi'}(v_2) < t$, by definition of $P_1'$ we know that $v_1$ is
	the first accept in $\pi'$, implying that no item before $\pos_{\pi'}(v_1) = \pos_\pi (v_2)$ is accepted by
	$\mathcal{A}_2$.
	In the case $\pos_{\pi'}(v_2) \geq t$, since $\pos_{\pi'}(v_1) \geq t$, the smallest rank in the sampling of $\pi'$ is 3 or
	greater. Therefore, $v_2$ is accepted if at most one item before $v_2$ is accepted. This holds for $\pi$, as $\pos_\pi(v_2) = \pos_{\pi'}(v_1)$.
\end{proof}

From $\card{P_1} = \card{P_1'} + \card{P_1 \setminus P_1'} = \card{P_2} + \card{P_1 \setminus P_1'}$
we get $p_1 = p_2 + \delta$, where $\delta = \card{P_1 \setminus P_1'} / n!$. 
That is, $\delta$ is the probability that a random permutation is element of $\card{P_1 \setminus P_1'}$. 
We analyze this event in \Cref{lemma:optimisticP1} using a similar counting argument as in the proof of \Cref{lemma:PrJthAccept}.

\begin{lemma}
	\label{lemma:optimisticP1}
	Let $\delta = \Pr{\pi \in P_1 \setminus P_1'}$ where $\pi$ is drawn uniformly at random from the set of all
	permutations and $P_1'$ is defined as in \Cref{lemma:P2SubsetP1}.
	It holds that
	$\delta =  \frac{t-1}{n} \frac{t-2}{n-1} \sum_{i=t}^{n-1} \frac{n-i}{(i-2)(i-1)}$.
\end{lemma}
\begin{proof}
	The set $P_1 \setminus P_1'$ contains exactly those permutations where
	$v_2$ is in the sampling and $\mathcal{A}_2$ accepts $v_1$ as the second item.
	Hence, $s_1=v_2$, and there exists an accepted item $v_x$ before $v_1$.
	
	Let $z \geq 3$ be the rank of $s_2=v_z$	and $x \in \intInterRightOpen{3}{z}$ be the rank of the first accepted element $v_x$.
	Let $\pi$ be drawn uniformly at random.
	Consider the following sequence of random events.
	\begin{enumerate}[(i)]
		\item $\pos(v_x) = i$ for $i \in \intInterRightOpen{t}{n}$
		\item $\pos(v_1) = \ell$ for $\ell \in \intInterLeftOpen{i}{n}$
		\item $\pos(v_2) \leq t-1$
		\item $\pos(s_2) \leq t-1$
		\item All items with positions in $[i-1] \setminus \{\pos(s_2),\pos(v_2) \}$ have ranks greater than $z$.
	\end{enumerate}
	The above conditions exactly characterize the event $\pi \in P_1 \setminus P_1'$ (see also \Cref{fig:P1Complement}):
	They ensure that the best two sampling elements are $s_1=v_2$ and $s_2$ and
	fix the positions of $v_x$ and $v_1$ to be $i$ and $\ell$, respectively.
	Further, all elements before position $i$ except from $s_1$ and $s_2$ must have ranks greater than $z$,
	such that in fact $s_1$ and $s_2$ are the best two elements in the sampling and no item before $v_x$
	is selected by $\mathcal{A}_2$.
	Note that we do not need an extra event ensuring that $v_1$ gets accepted if it comes
	after the first accept: Since the second accept must beat $s_1=v_2$, the only item with this property is $v_1$.
	
	\begin{figure}
		\centering
		\includegraphics[width=0.8\textwidth]{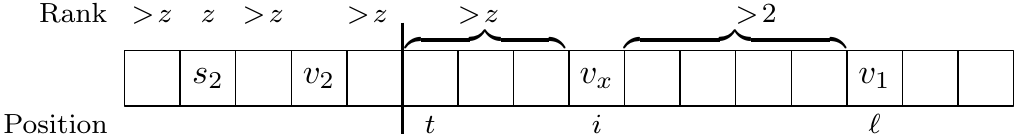}
		\caption{Event $\pi \in P_1 \setminus P_1'$ considered in \Cref{lemma:optimisticP1}.}
		\label{fig:P1Complement}
	\end{figure}
	
	The probability for the first four parts (i-iv) is $\beta := \frac{1}{n} \frac{1}{n-1} \frac{t-1}{n-2} \frac{t-2}{n-3}$.
	For event (v), we need the probability that the next $i-3$ items, drawn from the set of $n-4$ remaining items,
	all have rank greater than $z$. 
	As no item from $\{v_{z+1}, \ldots, v_n\}$ was drawn so far, $n-z$ items of rank greater than $z$ remain. 
	Therefore, by \Cref{fact:BlueBalls} the probability for step (v) is $h(n-4,n-z,i-3)$.
	By the law of total probability we obtain finally
	\begin{align}
	\delta &= \sum_{i=t}^{n-1} \sum_{\ell=i+1}^{n} \sum_{z=3}^{n} \sum_{x=3}^{z-1} \beta \cdot h(n-4,n-z,i-3)   \notag \\
	&= \beta \sum_{i=t}^{n-1} (n-i) \frac{1}{\binom{n-4}{i-3}} \sum_{z=3}^{n} (z-3) \binom{n-z}{i-3} \,. \label{eq:deltaSplitted}
	\end{align}
	The last term can be simplified further. First, we eliminate the sum over $z$ by applying (\ref{rule:sumOfProducts}):
	\begin{equation}
	\label{eq:deltaSumEliminated}
	\sum_{z=3}^{n} (z-3) \binom{n-z}{i-3}
%	= \sum_{z=0}^{n-3} z \binom{n-3-z}{i-3}
	= \sum_{z=0}^{n-3} \binom{z}{1} \binom{n-3-z}{i-3}
	= \binom{n-2}{i-1} \,.
	\end{equation}
	Using \Cref{eq:deltaSumEliminated} in \Cref{eq:deltaSplitted} yields
	\begin{equation*}
	\delta 
	= \beta \sum_{i=t}^{n-1} (n-i) \frac{\binom{n-2}{i-1}}{\binom{n-4}{i-3}} \\
	= \beta \sum_{i=t}^{n-1} (n-i) \frac{(n-3)(n-2)}{(i-2)(i-1)} \\
	\end{equation*}
	and the claim follows by resubstituting $\beta$.
\end{proof}

\subsection{Competitive Ratio}

Finally we can state the competitive ratio of \textsc{optimistic} in the case $k=2$.
Again, we consider the asymptotic setting where $n \to \infty$ and $t-1 = cn$ for some constant $c \in (0,1)$.

\begin{theorem}
	\textsc{optimistic} is $0.4168$-competitive for $k=2$ assuming $t-1 = cn$ for $c = 0.3521$.
\end{theorem}
\begin{proof}
	By the relation between \textsc{optimistic} and \textsc{single-ref} proven in \Cref{lemma:optimisticP2} and by \Cref{lemma:PrJthAccept}, we obtain $p_2 = \frac{t-1}{n} \sum_{i=t}^{n} \frac{1}{i-1}$. As proven in \Cref{lemma:dominatingItemsAsymptotic}, this term approaches
	$c \ln (1/c)$ in the asymptotic setting.
	Further, in \Cref{sec:optimisticP1} we showed $p_1 = p_2 + \delta$ where $\delta =  \frac{t-1}{n} \frac{t-2}{n-1} \sum_{i=t}^{n-1} \frac{n-i}{(i-2)(i-1)}$.
	The term $\sum_{i=t}^{n-1} \frac{n-i}{(i-2)(i-1)}$ is bounded asymptotically from above and below by
	$\frac{1}{c} - \ln \frac{1}{c} - 1$ (see \Cref{lemma:optimisticSumBound} of \ref{app:optimistic} for a proof).
	Further, $\frac{t-1}{n} \frac{t-2}{n-1} = c^2 - \frac{1-c}{n-1} = c^2 - o(1)$. 
	According to \Cref{lemma:compRatio}, the competitive ratio is
	\[
	\frac{1}{2} \left(p_1 + p_2 \right)
	= \frac{1}{2} \left(p_2 + \delta + p_2 \right)
	= c \ln \frac{1}{c} + \frac{c^2}{2} \left( \frac{1}{c} - \ln \frac{1}{c} - 1\right) \,.
	\]
	The optimal choice for $c$ is around $c^*=0.3521 < 1/e$, which gives a competitive ratio of $0.4168$.
\end{proof}

\section{Conclusion and Future Work}
\label{sec:conclusion}
In this work, we investigated two algorithms for the $k$-secretary problem with a focus on small values for $k \geq 2$.
We introduced and analyzed the algorithm \textsc{single-ref}. For any value of $k$, the competitive ratio of \textsc{single-ref} can be obtained by numerical optimization. Further, we provided a tight analysis of the \textsc{optimistic} algorithm \cite{DBLP:conf/approx/BabaioffIKK07} in the case $k=2$.

We see various directions of future work. 
For \textsc{single-ref}, it remains to find the right dependency between the parameters 
$r$, $c$, and $k$ in general and, if possible, to find a closed formula for the competitive ratio for any value of $k$. 
\textsc{optimistic} seems a promising and elegant algorithm, however no tight analysis for general $k \geq 3$ is known so far. 
For $k=2$, we identified a key property in \Cref{lemma:optimisticP2}. Similar properties may hold in the general case.

\bibliography{literature}

\newpage
\appendix

\section{Optimal Parameters for SINGLE-REF}
\label{app:single-refTable}

\begin{table}[h!]
	\centering
	\caption{Optimal parameters and corresponding competitive ratios (c.r.) of \textsc{single-ref} for $k \in \intInter{1}{100}$.
		For readability, the numeric values are truncated after the fourth decimal place.}
	\label{tab:comeptitiveRatios20Full}	
	\vspace*{8pt}
	\begin{tabular}{c@{\hskip 1\tabcolsep} c@{\hskip 1.5\tabcolsep} c@{\hskip 1.5\tabcolsep} c|} 
		\toprule
		$k$ & $r$ & $c$ & c.r.\\ 
		\midrule 
		1 & 1 & 0.3678 & 0.3678 \\ 
		2 & 1 & 0.2545 & 0.4119 \\ 
		3 & 2 & 0.3475 & 0.4449 \\ 
		4 & 2 & 0.2928 & 0.4785 \\ 
		5 & 2 & 0.2525 & 0.4999 \\ 
		6 & 2 & 0.2217 & 0.5148 \\ 
		7 & 3 & 0.2800 & 0.5308 \\ 
		8 & 3 & 0.2549 & 0.5453 \\ 
		9 & 3 & 0.2338 & 0.5567 \\ 
		10 & 3 & 0.2159 & 0.5660 \\ 
		11 & 4 & 0.2570 & 0.5740 \\ 
		12 & 4 & 0.2410 & 0.5834 \\ 
		13 & 4 & 0.2267 & 0.5914 \\ 
		14 & 4 & 0.2140 & 0.5983 \\ 
		15 & 4 & 0.2026 & 0.6043 \\ 
		16 & 4 & 0.1924 & 0.6096 \\ 
		17 & 5 & 0.2231 & 0.6155 \\ 
		18 & 5 & 0.2133 & 0.6211 \\ 
		19 & 5 & 0.2042 & 0.6261 \\ 
		20 & 5 & 0.1959 & 0.6306 \\ 
		21 & 5 & 0.1882 & 0.6347 \\ 
		22 & 5 & 0.1811 & 0.6384 \\ 
		23 & 6 & 0.2054 & 0.6426 \\ 
		24 & 6 & 0.1985 & 0.6465 \\ 
		25 & 6 & 0.1919 & 0.6502 \\ 
		26 & 6 & 0.1858 & 0.6535 \\ 
		27 & 6 & 0.1800 & 0.6566 \\ 
		28 & 6 & 0.1746 & 0.6595 \\ 
		29 & 7 & 0.1947 & 0.6625 \\ 
		30 & 7 & 0.1893 & 0.6655 \\ 
		31 & 7 & 0.1842 & 0.6684 \\ 
		32 & 7 & 0.1793 & 0.6711 \\ 
		33 & 7 & 0.1747 & 0.6736 \\ 
		34 & 7 & 0.1703 & 0.6760 \\ 
%		\bottomrule
	\end{tabular}
%	\hspace*{-15pt}
	\begin{tabular}{c@{\hskip 1\tabcolsep} c@{\hskip 1.5\tabcolsep} c@{\hskip 1.5\tabcolsep} c|} 
		\toprule
		$k$ & $r$ & $c$ & c.r.\\ 
		\midrule 
		35 & 7 & 0.1662 & 0.6782 \\ 
		36 & 8 & 0.1830 & 0.6805 \\ 
		37 & 8 & 0.1788 & 0.6829 \\ 
		38 & 8 & 0.1748 & 0.6851 \\ 
		39 & 8 & 0.1710 & 0.6873 \\ 
		40 & 8 & 0.1673 & 0.6893 \\ 
		41 & 8 & 0.1638 & 0.6912 \\ 
		42 & 8 & 0.1605 & 0.6930 \\ 
		43 & 9 & 0.1750 & 0.6948 \\ 
		44 & 9 & 0.1716 & 0.6968 \\ 
		45 & 9 & 0.1683 & 0.6986 \\ 
		46 & 9 & 0.1651 & 0.7004 \\ 
		47 & 9 & 0.1621 & 0.7021 \\ 
		48 & 9 & 0.1592 & 0.7037 \\ 
		49 & 9 & 0.1563 & 0.7052 \\ 
		50 & 9 & 0.1536 & 0.7067 \\ 		
		51 & 10 & 0.1662 & 0.7082 \\ 
		52 & 10 & 0.1635 & 0.7098 \\ 
		53 & 10 & 0.1608 & 0.7113 \\ 
		54 & 10 & 0.1582 & 0.7127 \\ 
		55 & 10 & 0.1557 & 0.7141 \\ 
		56 & 10 & 0.1532 & 0.7155 \\ 
		57 & 10 & 0.1509 & 0.7168 \\ 
		58 & 10 & 0.1486 & 0.7180 \\ 
		59 & 11 & 0.1597 & 0.7193 \\ 
		60 & 11 & 0.1574 & 0.7206 \\ 
		61 & 11 & 0.1551 & 0.7219 \\ 
		62 & 11 & 0.1529 & 0.7231 \\ 
		63 & 11 & 0.1508 & 0.7243 \\ 
		64 & 11 & 0.1487 & 0.7255 \\ 
		65 & 11 & 0.1467 & 0.7266 \\ 
		66 & 11 & 0.1447 & 0.7277 \\ 
		67 & 11 & 0.1428 & 0.7287 \\ 
		68 & 12 & 0.1527 & 0.7298 \\ 
	%	\bottomrule
	\end{tabular}		
%	\hspace*{-15pt}
	\begin{tabular}{c@{\hskip 1\tabcolsep} c@{\hskip 1.5\tabcolsep} c@{\hskip 1.5\tabcolsep} c} 
		\toprule
		$k$ & $r$ & $c$ & c.r.\\ 
		\midrule 
		69 & 12 & 0.1508 & 0.7309 \\ 
		70 & 12 & 0.1489 & 0.7320 \\ 
		71 & 12 & 0.1470 & 0.7330 \\ 
		72 & 12 & 0.1452 & 0.7340 \\ 
		73 & 12 & 0.1434 & 0.7350 \\ 
		74 & 12 & 0.1417 & 0.7360 \\ 
		75 & 12 & 0.1400 & 0.7369 \\ 
		76 & 12 & 0.1384 & 0.7378 \\ 
		77 & 13 & 0.1473 & 0.7387 \\ 
		78 & 13 & 0.1456 & 0.7397 \\ 
		79 & 13 & 0.1440 & 0.7406 \\ 
		80 & 13 & 0.1424 & 0.7415 \\ 
		81 & 13 & 0.1408 & 0.7424 \\ 
		82 & 13 & 0.1393 & 0.7433 \\ 
		83 & 13 & 0.1378 & 0.7441 \\ 
		84 & 13 & 0.1363 & 0.7449 \\ 
		85 & 13 & 0.1349 & 0.7457 \\ 
		86 & 14 & 0.1429 & 0.7465 \\ 
		87 & 14 & 0.1415 & 0.7473 \\ 
		88 & 14 & 0.1400 & 0.7482 \\ 
		89 & 14 & 0.1386 & 0.7490 \\ 
		90 & 14 & 0.1372 & 0.7497 \\ 
		91 & 14 & 0.1359 & 0.7505 \\ 
		92 & 14 & 0.1346 & 0.7512 \\ 
		93 & 14 & 0.1333 & 0.7520 \\ 
		94 & 14 & 0.1320 & 0.7527 \\ 
		95 & 14 & 0.1307 & 0.7534 \\ 
		96 & 15 & 0.1381 & 0.7541 \\ 
		97 & 15 & 0.1368 & 0.7548 \\ 
		98 & 15 & 0.1356 & 0.7555 \\ 
		99 & 15 & 0.1343 & 0.7562 \\ 
		100 & 15 & 0.1331 & 0.7569 \\ 
		& & & \\		
		& & & \\				
	%	\bottomrule
	\end{tabular}			
\end{table}

\newpage
\section{Omitted Proofs for SINGLE-REF}
\label{app:single-ref}

\begin{lemma}
	\label{lemma:fMonotoneSingleMaximum}
	Let $f \colon \RR_{\geq 0} \to \RR_{\geq 0}$ with $f(i) = i^j / (i+y)^{r+j}$ and constants $j \geq 0$, $r \geq 1$, and $y > 0$. Define $z= \frac{jy}{r}$.
	The function $f$ has the following properties:
	\begin{enumerate}[(A)]
		\item $f$ has a global maximum point at $z$,
		\item $f$ is monotonically increasing on $[0,z]$ and monotonically decreasing on $[z,\infty)$.
	\end{enumerate}
\end{lemma}
\begin{proof}
	Let $g(i) = i^j$ and $h(i) = (i+y)^{r+j}$, thus $f(i)=g(i)/h(i)$. For the first derivative of $f$ we obtain $f'(i) = \frac{g'(i) h(i) - g(i) h'(i)}{h(i)^2}$.
	Since $h(i)^2$ is non-negative for all $i \geq 0$, we have
	\begin{align*}
	f'(i) \geq 0
	&\Leftrightarrow 
	g'(i) h(i) \geq g(i) h'(i)  \\
	&\Leftrightarrow~ 	j i^{j-1} (i+y)^{r+j} \geq i^j (r+j) (i+y)^{r+j-1} \\
	&\Leftrightarrow~ 	j (i+y) \geq i (r+j)  \\
	&\Leftrightarrow 	z \geq i \,.
	\end{align*}
	Hence, $f$ is monotonically increasing on $[0,z]$ and monotonically decreasing on $[z,\infty)$.
	An analogous calculus shows $f'(i)=0$ if and only if $i=z$. Therefore, $z$ is a global maximum point.
\end{proof}

\begin{lemma}
	\label{lemma:fAntiderivativeR1}
	Let $f \colon \RR \to \RR$ with $f(i) = i^j / (i+y)^{r+j}$ and constants $j \geq 0$, $r=1$, and $y > 0$.
	The following function $F$ fulfills $F'(i) = f(i)$:
	\[
	F(i) = \ln(i+y) - \sum_{\ell=1}^{j} \frac{\beta_\ell}{\ell} \cdot \left( \frac{y}{i+y}\right)^\ell \,,
	\]
	where $\beta_\ell = (-1)^\ell \binom{j}{\ell}$ for $1 \leq \ell \leq j$.
\end{lemma}
\begin{proof}
	We have
	\begin{align*}
	F'(i)
	&= \frac{1}{i+y} - \sum_{\ell=1}^{j} \frac{\beta_\ell}{\ell} \cdot \ell \cdot \left( \frac{y}{i+y} \right)^{\ell - 1} \cdot \left( - \frac{y}{(i+y)^2}\right) \\
	&= \frac{1}{i+y} + \sum_{\ell=1}^{j} \beta_\ell \cdot \frac{y^\ell}{(i+y)^{\ell+1}} \\
	&= \sum_{\ell=0}^{j} \beta_\ell \cdot \frac{y^\ell}{(i+y)^{\ell+1}} \\	
	&= \sum_{\ell=0}^{j} (-1)^\ell \binom{j}{\ell} \cdot \frac{y^\ell}{(i+y)^{\ell+1}} \\
	&= \frac{1}{(i+y)^{j+1}} \cdot \sum_{\ell=0}^{j} \binom{j}{\ell} \cdot (-y)^\ell \cdot (i+y)^{j-\ell} \\
	&= \frac{i^j}{(i+y)^{j+1}} \,,
	\end{align*}
	where the last inequality follows from the binomial theorem: For all $a, b, n \in \mathds{Z}$ it holds that $(a+b)^n = \sum_{k=0}^{n} \binom{n}{k} a^{k} b^{n-k}$.
\end{proof}

\begin{lemma}
	\label{lemma:fAntiderivativeR2}
	Let $f \colon \RR \to \RR$ with $f(i) = i^j / (i+y)^{r+j}$ and constants $j \geq 0$, $r \geq 2$, and $y > 0$.
	The following function $F$ fulfills $F'(i) = f(i)$:
	\[
	F(i) = - \frac{\sum_{\ell=0}^{j} \alpha_\ell  i^{j-\ell} y^\ell}{\alpha_0 (r-1) (i+y)^{r+j-1}} \,,
	\]
	where $\alpha_\ell = \binom{j+r-1}{\ell+r-1}$ for $0 \leq \ell \leq j$.
\end{lemma}
\begin{proof}
	Let $G(i) = - \sum_{\ell=0}^{j} \alpha_\ell  \cdot i^{j-\ell} \cdot y^\ell$ and 
	$H(i) = \alpha_0 \cdot (r-1) \cdot (i+y)^{r+j-1}$ be the numerator and denominator of $F(i)$, respectively.
	We derive the first derivatives $G'(i) = - \sum_{\ell=0}^{j} \alpha_\ell  \cdot i^{j-\ell-1} \cdot y^\ell \cdot (j-\ell)$ and 
	$H'(i) = \alpha_0 \cdot (r-1) \cdot (r+j-1) \cdot (i+y)^{r+j-2} = H(i) \cdot \frac{r+j-1}{i+y}$. 
	Therefore,
	\begin{align*}
		F'(i) 
		&= \frac{G'(i) \cdot H(i) - G(i) \cdot H'(i)}{H(i)^2} \\
		&= \frac{G'(i) - G(i) \cdot \frac{r+j-1}{i+y}}{H(i)}  \\
		&= \frac{G'(i) \cdot (i+y) - G(i) \cdot (r+j-1)}{\alpha_0 \cdot (r-1) \cdot (i+y)^{r+j}} \,.
	\end{align*}
	Hence, the claim follows if we can show
	\begin{equation}
	G'(i) \cdot (i+y) - G(i) \cdot (r+j-1) = i^j \cdot \alpha_0 \cdot (r-1) \,.
	\label{eq:antiDerivativeHelper}	
	\end{equation}
	To show \Cref{eq:antiDerivativeHelper}, we observe
	\begin{align*}
	& \quad G'(i) \cdot (i+y) - G(i) \cdot (r+j-1) \\
	&= - \left( \sum_{\ell=0}^{j} \alpha_\ell  \cdot i^{j-\ell-1} \cdot y^\ell \cdot (j-\ell) \right) \cdot (i+y)
	+ \left( \sum_{\ell=0}^{j} \alpha_\ell  \cdot i^{j-\ell} \cdot y^\ell \right) \cdot (r+j-1) \\
	&= - \left( \sum_{\ell=0}^{j} \alpha_\ell  \cdot i^{j-\ell} \cdot y^\ell \cdot (j-\ell) \right)
	- \left( \sum_{\ell=0}^{j} \alpha_\ell  \cdot i^{j-\ell-1} \cdot y^{\ell+1} \cdot (j-\ell) \right)  \\
	&   ~~~ + \left( \sum_{\ell=0}^{j} \alpha_\ell  \cdot i^{j-\ell} \cdot y^\ell \cdot (r+j-1) \right) \\
	&= \left( \sum_{\ell=0}^{j} \alpha_\ell  \cdot i^{j-\ell} \cdot y^\ell \cdot (r-1+\ell) \right)
	- \left( \sum_{\ell=0}^{j} \alpha_\ell  \cdot i^{j-\ell-1} \cdot y^{\ell+1} \cdot (j-\ell) \right) \,.
	\end{align*}
	Let $S_1$ and $S_2$ be the first and the second sum in the last expression, respectively.
	Note that $S_1$ can be rewritten as follows
	\begin{align*}
	S_1 
	&= \alpha_0 \cdot i^j \cdot (r-1) + \sum_{\ell=1}^{j} \alpha_\ell  \cdot i^{j-\ell} \cdot y^\ell \cdot (r-1+\ell) \\
	&= \alpha_0 \cdot i^j \cdot (r-1) + \sum_{\ell=1}^{j} \alpha_{\ell-1}  \cdot i^{j-\ell} \cdot y^\ell \cdot (j-\ell+1) \\
	&= \alpha_0 \cdot i^j \cdot (r-1) + S_2 \,,
	\end{align*}
	where we used the fact that $\frac{\alpha_\ell}{\alpha_{\ell-1}} = \frac{j-\ell+1}{\ell+r-1}$ for $1 \leq \ell \leq j$ for the second equality.
	This proves Equation~(\ref{eq:antiDerivativeHelper}) and concludes the proof of the lemma.
\end{proof}

\section{Omitted Proofs for OPTIMISTIC}
\label{app:optimistic}

\begin{lemma}
	\label{lemma:optimisticSumBound}
	Assuming $t-1=cn$ for $c \in (0,1)$, it holds that
	\[
	\frac{1}{c} - \ln \frac{1}{c} - 1 - o(1) 
	\leq \sum_{i=t}^{n-1} \frac{n-i}{(i-2)(i-1)} 
	\leq \frac{1}{c} - \ln \frac{1}{c} - 1 + o(1) \,.
	\]
\end{lemma}
\begin{proof}
	The lower and upper bounds follow basically from \Cref{lemma:approxSumByIntegral}A.
	For the lower bound, note that $(n-i)/i^2$ decreases monotonically in $i$. Therefore,
	\begin{multline*}
	\sum_{i=t}^{n-1} \frac{n-i}{(i-2)(i-1)} 
	> \sum_{i=t}^{n-1} \frac{n-i}{i^2} 
	\geq \int_{t}^n \frac{n-i}{i^2} \intD{i}
	= \ln \frac{t}{n} + \frac{n}{t} - 1 \\
	> \ln \frac{t-1}{n} + \frac{n}{t-1} \cdot \frac{t-1}{t} - 1
	= \frac{1}{c} - \ln \frac{1}{c} - 1 - \underbrace{\frac{1}{ct}}_{=o(1)} \,.
	\end{multline*}
	The upper bound follows likewise. Observe that
	\[
	\sum_{i=t}^{n-1} \frac{n-i}{(i-2)(i-1)} 
	< \sum_{i=t}^{n} \frac{n-i}{(i-2)^2}
	< \sum_{i=t-2}^{n} \frac{n-i}{i^2}
	= \left(\sum_{i=t}^{n} \frac{n-i}{i^2}\right) + \xi \,,
	\]
	where 
	\[
	\xi = \frac{n-(t-2)}{(t-2)^2} + \frac{n-(t-1)}{(t-1)^2} = o(1) \,.
	\]
	Since $\frac{n+2-i}{i^2}$ decreases monotonically in $i$, we obtain
	\begin{multline*}
	\left(\sum_{i=t}^{n} \frac{n-i}{i^2}\right) + \xi
	\leq \left(\int_{t-1}^{n} \frac{n-i}{i^2} \intD{i}\right) + \xi 
	= \ln \frac{t-1}{n} + \frac{n}{t-1} - 1 + \xi \\
	= \frac{1}{c} - \ln \frac{1}{c} - 1 + \xi \,.
%	\qedhere
	\end{multline*}
\end{proof}

\end{document}